\documentclass[prb,aps,epsf]{revtex4}

\usepackage[dvips]{graphicx}                              
\begin{document}

\title{Field Theory of Anisotropic Quantum Hall Gas: Metrology
and a Novel Quantum Hall Regime
} 

\author{K. Ishikawa, T. Aoyama, Y. Ishizuka, and N. Maeda}

\address{Department of Physics, Hokkaido University, 
Sapporo, 060-0810, Japan
}

\begin{abstract}
The von Neumann lattice representation is a convenient representation 
for studying several intriguing physics of quantum Hall systems. 
In this formalism, electrons are mapped to lattice fermions. 
A topological invariant expression of the Hall conductance 
is derived and is used for the proof of the integer quantum Hall effect in 
the realistic situation. Anisotropic quantum Hall gas is investigated 
based on the Hartree-Fock approximation  in the same formalism. 
Thermodynamic properties, transport properties, and unusual response 
under external modulations are found. Implications for the integer 
quantum Hall effect in the finite systems are also studied and a new 
quantum Hall regime with non-zero longitudinal resistance is shown to exist. 

\end{abstract}

\maketitle

\section{ Introduction }

A basic physical law is described by a basic equation which includes 
usually a fundamental physical constant. The following constants are such 
parameters and play important roles in modern physical science. 
\begin{table}[h]
\caption{Three fundamental physical constants.}
{\begin{tabular}{@{}ccc@{}}\toprule  
Physical Constant & Symbol & Physical law \\ \colrule
electric charge   &   $e$     & unit of charge \\
light velocity    &   $c$     & special relativity\\
Plank constant    &   $h$     & quantum mechanics \\ \botrule
\end{tabular}}
\end{table}
These constants make the physics of matter, physics of space and time, 
and physics of microscopic world different completely from the 
classical world. It is important to measure them precisely. 

First constant, $e$, is  the electric charge 
of the electron, and is the unit of charge. Charge of any object is 
known to be integer multiple of $e$. 
Millikan found the fact that the charge is not continuous but is quantized. 
Atom is composed of a nucleus and electrons. Nucleus has 
positively quantized charge. So by combining nucleus with suitable number of 
electrons, the system becomes neutral. The fact that the charge is quantized 
is important for the neutrality of matter. 

Concerning second constant, $c$, the fact that the light velocity is finite 
was known for a long time. It was discovered that the light velocity is 
universal in the last century. Velocity of light which is emitted from 
moving matter is $c$ and the velocity of light that is measured by moving 
apparatus is also $c$. This constant light velocity plays critical 
roles for special relativity. 
 
The third constant is Plank constant, $h$, which has the unit of action and 
plays the important role in microscopic world. When the light of a 
frequency $\nu$ interacts with microscopic systems, 
the light has an energy, $h \nu$. The light behaves like a particle in 
microscopic world. The electron which is a particle in classical 
mechanics becomes a wave of wave length $\lambda$. The Plank 
constant determines the wave length, $\lambda$, of the electron of 
definite momentum, $p$, as $\lambda=h/p$. Particle is compatible with wave 
in quantum mechanics.

Physics which is connected with all these physical constants, $e$, 
$c$, and $h$ may reveal fundamental dynamical principle of nature. 
One area of physics which is connected with all of them is quantum 
electrodynamics (QED). QED has a long history and has been well established 
by now. 
However QED is still an important topic of fundamental physics and may shed 
a new light for understanding our nature. A unique parameter of QED is the 
fine structure constant, ${\alpha}$, defined by 
\begin{equation}
{\alpha}={e^2 \over 2 \varepsilon_{0}ch},
\end{equation}
where ${\epsilon}_0$ is the dielectric constant of vacuum. Another area 
of physics which is connected with the fine structure constant is the 
quantum Hall effect (QHE). The QHE is the phenomenon  in semiconductors and 
is connected with fundamental science.

\subsection{Fine structure constant of Quantum Electrodynamics (QED)}

QED is the quantum field theory of electron and 
photon which satisfies requirement of relativistic invariance, and the 
foundation of electromagnetic force is given from QED. In QED, electron 
field describes the electron and satisfies relativistic Dirac equation. 
Dirac equation is the relativistic extension of Schr{\"o}dinger equation and 
has a form of first order differential equation with Hermitian Hamiltonian. 
Hence probability is preserved as is in ordinary quantum mechanics. 
Peculiar feature of Dirac equation is that the equation has 
negative energy solutions in addition to positive energy solutions. 
Transition of electron in positive energy state to that in negative energy 
state by emitting a photon should have a finite probability if the 
corresponding negative energy state is empty. 
Thus a particle state is unstable. If all the negative energy states are 
occupied in the vacuum and only one electron is allowed in one state, then 
the stability is ensured. 
From this idea, Dirac is lead to introduce Dirac sea. 
Namely the vacuum is not empty but is full of electrons of negative energy 
states. All the negative energy states are filled with electrons in the 
vacuum. Then the transition of electron of positive energy state into the 
electron of negative energy state and photon does not occur. 
The stability is ensured. Because vacuum is full of negative energy 
electrons, a hole of Dirac sea behaves like a particle of 
a positive charge. This particle is anti-particle of electron and is called 
positron.

Photon is described by Maxwell equations that is invariant under Lorenz 
transformation and has no negative energy solution. But each mode of a 
definite wave vector is equivalent to a harmonic oscillator and in quantum 
theory each mode has finite energy due to zero point oscillation. Thus the 
ground state, vacuum, of QED is not actually real empty state but 
has rich ingredients of negative energy electrons and zero point oscillation 
of photons. Physical phenomena which are generated from these dynamical 
freedoms present characteristic features of nature in the microscopic 
world. 

One example of physical quantity which is generated from the vacuum 
fluctuation is seen in an isolated electron in vacuum. One electron has 
influences from these rich dynamical effects of vacuum. Mass and electric 
charge are those physical quantities that are affected from these effects. 
If we knew both of bare values and the real values of these quantities, 
it would have been possible to observe the interaction effect. However in the 
real world the electron always interacts with photon and has an influence 
from the electrons in negative energy states. Bare values of mass and 
charge are unobservable. Therefore mass and charge are not appropriate to 
see the vacuum fluctuation effects. The anomalous magnetic moment of the 
electron is one of the physical quantities that is due to vacuum fluctuation 
effect. In fact,the anomalous magnetic moment of the 
electron gives an important information of the vacuum. 

The electron's anomalous magnetic moment is given by, 
\begin{equation}
\mu=g{eh \over 2m_{e}},
\end{equation}
where $m_e$ is the mass of electron. Here the gyro-magnetic ratio, $g$, is 
determined to 2 from Dirac equation, if there is no correction from 
vacuum fluctuation. The real value of $g$ is deviated from 2 slightly and 
the $g-2$ is due to vacuum fluctuation. The $g-2$ is computed theoretically 
as a power series of the fine structure constant in QED and is measured 
experimentally with extraordinary precision.
 
We compare the theoretical value of $g-2$ with the experimental 
value. From higher order perturbative calculation, Kinoshita\cite{kinoshita} 
has obtained, 
\begin{eqnarray}
g-2 &=& {\alpha \over 2 \pi}-0.328478965({\alpha \over \pi})^2  \nonumber \\
    & &+1.17562(56)({\alpha \over \pi})^3+(-)1.472(152)({\alpha \over \pi})^4 
+4.46 {\times}10^{-12}.
\end{eqnarray}
Here, $\alpha$ is the fine structure constant and the last term in the 
right-hand side of the above equation is the estimates from weak interaction. 
Experiment by Van Dyck et al.\cite{van dyck} uses an electron in Penning 
trap for measuring electron's magnetic moment. The value of $g-2$ is given as, 
\begin{equation}
g-2=1159652188.4(4.3)\times 10^{-12}.
\end{equation}
Comparing the experimental value of $g-2$ with the theoretical value, 
we have the ${\alpha}^{-1}$ from $g-2$, 
\begin{equation}
{\alpha}_{g-2}^{-1}=137.03599976(50) (3.7ppb).
\end{equation}
On the other hand, the fine structure constant is determined directly 
by quantum Hall effect (QHE), 
\begin{equation}
{\alpha}_{\rm QHE}^{-1}=137.0360037(33) (0.024ppm).
\end{equation}

The value of fine structure constant obtained from the electron's magnetic 
moment is slightly different from the value that has been 
obtained from QHE. Actually QHE gives the most accurate value of direct 
measurement of ${\alpha}$.  Now this difference is too small to 
conclude something definite, but in future it may become possible. So it 
is important to clarify if QHE gives the precise value of ${\alpha}$. 

QHE's are physical phenomena in two dimensional electrons with a strong 
magnetic field realized in semiconductors and are connected also with the 
above fundamental constants. Integer quantum Hall effect (IQHE) and 
fractional quantum Hall effect (FQHE) were found nearly twenty years ago 
and many other interesting phenomena have been found since then. 
Study of new phase of QHE discovered recently and its implication 
to the precise determination of the fine structure 
constant are presented in the present paper.

\subsection{ Quantum Hall effects}

In interface of special semiconductors such as GaAs-AlGaAs heterojunctions 
or Silicon MOSFET, electron's motion are restricted to a plane. 
When strong perpendicular magnetic field is applied, electron's two 
dimensional motion is frozen due to magnetic field. As was solved by Landau 
first\cite{landau}, electron's energy becomes discrete and degeneracy of 
each energy level per area, density $\rho_0$, is proportional to the 
magnetic field, 
\begin{eqnarray}
E_l&=&{\hbar eB \over m}(l+{1 \over 2}), \nonumber \\
\rho_0&=&{eB \over 2\pi\hbar},        
\end{eqnarray}
where $B$ is the magnetic field and $m$ is the electron's mass and $l$ 
is a non-negative integer. In  semiconductors there exist disorders which 
make these electrons been bound. Electrons are localized around disorders 
and do not carry electric current if their energies are away from one of 
the above discrete values. Conversely electrons with the energy 
of the above discrete values are not localized but are extended. 
Because the electrons in the extended states carry electric current, 
Hall conductance changes its value with a Fermi energy if the Fermi energy 
is in the energy region of extended state. If the Fermi energy is in the 
the localized state region, the Hall conductance stays to constant value. 
The Hall conductance at plateau takes very special quantized value. 
This is the IQHE\cite{klitzing}.  In high mobility samples, on the other 
hand, interactions make electron system form new condensed 
states\cite{tsui,laughlin}. From these effects Hall resistance and 
longitudinal resistance are varied as shown in Fig.~1(a). Many structures 
are seen. When the longitudinal resistance vanishes the Hall resistance is 
quantized. Vanishing longitudinal resistance means that the ground state at 
plateau has energy gap. The ground states of the IQHE and the FQHE have 
the energy gap and are incompressible. In the former system cyclotron 
energy gap is the origin of gap and in the latter system interactions 
among electrons is the origin of energy gap. 

%
%
\begin{figure}[th]
\vspace*{5cm}
\caption{Experiments of the Hall resistance and longitudinal resistance. 
(a) In IQHE and FQHE, the Hall resistance is quantized as 
$({e^2 \over h}n )^{-1}$ or as  $({e^2 \over h}{p \over q})^{-1}$ 
and longitudinal resistance vanishes\cite{stormer}. Compressible gas 
phases are seen at $\nu={1 \over 2}$, $\nu={3 \over 2}$ and at higher 
Landau levels. 
(b) Huge anisotropic resistances are seen at around half filled higher 
Landau levels\cite{eisenstein}. The solid line shows the resistance for 
the current in vertical direction and the dashed line show the resistance 
for the current in horizontal direction.}
\end{figure}

At plateau Hall conductance is given as 
\begin{equation}
\sigma_{xy}={e^2 \over h} n,
\end{equation}
around $\nu =n({\rm integer})$, or as 
\begin{equation}
\sigma_{xy}={e^2 \over h} {q \over p},
\end{equation}
around $\nu ={q \over p}$, where $\nu =\rho/\rho_0$ and $p$ and $q$ are
coprime integers. Here $\rho$ is the electron's density. 
The former is the IQHE and the latter is FQHE. The value is quantized 
exactly despite the fact that the system has disorders and interactions. 
There is a special reason why quantization is exact in these systems. 
The Hall conductance is a topological invariant of the mapping defined by 
propagator of the electrons in Landau levels. This is a reason why the Hall 
conductance is quantized in the system of interactions and impurities and 
is given in the next section.   

In addition to these quantum Hall states, compressible quantum Hall states 
have been seen at around $\nu =1/2$ and  $\nu =3/2$ and higher Landau 
levels. The compressible states at $\nu =1/2$ and  $\nu =3/2$ are 
isotropic and expressed by composite Fermion theory\cite{jain}. 
A new exotic phenomenon at higher Landau levels, where longitudinal 
resistance $\rho_{xx}$ does not vanish and the Hall resistance $\rho_{xy}$ 
is not quantized, was found recently\cite{eisenstein,stripe experiment} as 
in Fig.~1(b). Extremely anisotropic resistance, where $\rho_{xx}$ is very 
different from  $\rho_{yy}$ was found. This phenomenon is caused by new 
many-body state which  has no energy gap and behaves like anisotropic gas. 
Although bare kinetic energy was frozen due to magnetic field, 
effective kinetic energy is generated by interactions in the Hall gas phase. 

We concentrate to study on the anisotropic quantum Hall gas phase and 
its implications to the metrology of QHE. In the anisotropic quantum Hall 
gas, one-particle kinetic energy which violates orientational and 
translational symmetries is generated. To study them we apply a Hartree-Fock 
mean field theory using a special representation which is symmetric in two 
directions. This representation, the von Neumann lattice representation, is 
convenient to study field theory of the Quantum Hall systems. 
Spectrum of Nambu-Goldstone mode and response of the system under external 
density modulation are found. Hall gas has several peculiar thermodynamic 
properties which are connected with spontaneous generation of the 
kinetic energy. As will be seen later in the present work, these properties 
guarantee a stability and precision of the IQHE in the realistic systems. 
It is shown that the precise determination of the fine structure constant 
is possible from QHE. 

This paper is organized in the following manner. 

 In Section 2, the von Neumann lattice representation of non-commutative 
coordinates and formulation of field theory of quantum Hall systems, i.e., 
systems of two dimensional electrons in the magnetic field, based on the 
von Neumann lattice representation are presented. Hamiltonian and current 
operators are expressed with the electrons in the von Neumann lattice 
representation and Ward-Takahashi identity are derived. Using these 
relations low energy theorem on the Hall conductance is derived. The Hall 
conductance is written as a special topological invariant of the mapping 
defined by the propagator of Landau levels and agrees with integer multiple 
of ${e^2 \over h}$ if certain boundary condition is satisfied. 
Values of the topological invariant in the system of disorders, 
interactions, and periodic potentials are found.  

In Section 3, a mean field theory of anisotropic Hall gas (stripe) is 
presented. A one-particle kinetic energy is generated by interactions and 
a Fermi surface of the Hall gas in the von Neumann lattice is identified. 
Thermodynamic properties and transport properties are reviewed. 
Several spatial symmetries such as translations and a rotation are broken 
in the present mean field. Nambu-Goldstone zero energy excitations connected 
with these symmetry breaking are implied by Goldstone theorem. 
The excitations are found to have a peculiar low energy behavior based on 
the single mode approximation. Easy direction, in which the resistance is 
extremely smaller than the other direction, becomes orthogonal to the 
external density modulation of long wave length. 
This behavior is opposite to that of ordinary charge density wave of 
zero-magnetic field.
                                                                          
In Section 4, implications of unusual thermodynamic properties of the 
Hall gas are summarized. Due to interactions or disorders, strip of 
compressible states are formed. 

In Section 5, implications of stripe and a strip of the Hall gas 
to IQHE in realistic systems are presented. It is pointed out that the 
precision and stability of the IQHE in the realistic systems are derived 
from unusual properties of the Hall gas. As the external current increases, 
several transport regimes such as quantum Hall regime (QHR), dissipative QHR, 
collapse, and breakdown appear. 

Summary is given in Section 6.

\section{The von Neumann lattice representation and topological expression 
of the  Hall conductance}
 
Under the perpendicular magnetic field, the electron's motion in the 
planar space becomes different from that of the zero magnetic field. The 
representation which preserves characteristic features of the system 
such as discrete energy with finite degeneracy that is proportional to 
the magnetic field, magnetic translational invariance, and non-commutative 
guiding center coordinates is convenient to develop field theories. 
The von Neumann lattice representation\cite{von neumann1,von neumann2} is 
such representation that is symmetric in two directions and is invariant 
under lattice translations. Universal identities are used to give a proof 
of the IQHE.

\subsection{ The von Neumann lattice representation}

One-body Hamiltonian for the two-dimensional electrons in the perpendicular 
magnetic fields is given as, 
\begin{eqnarray}
H_0&=&{({\bf p}+e{\bf A})^2 \over 2m}, \nonumber \\
& &{\partial_x} A_y -{\partial_y} A_x=B.
\end{eqnarray}
Since the velocity is connected with the momentum,
\begin{eqnarray}
v_x&=&{(p+eA)_x \over m},  \nonumber \\
v_y&=&{(p+eA)_y \over m},
\end{eqnarray}
components of the velocity satisfy 
\begin{equation}
[v_x,v_y]=-{i \hbar eB \over m^2}.
\end{equation}
The above Hamiltonian has a form that is equivalent to the Hamiltonian 
without magnetic field if it is written by velocity. Solutions of classical 
equation of motion with integration constants, $(X,Y)$  are,
\begin{eqnarray}
x &=&  {v_y \over \omega_c}+X,\nonumber \\
y &=& -{v_x \over \omega_c}+Y,
\end{eqnarray}
where ${\omega}_c=eB/m$. This is a circular motion around  the center
$(X,Y)$ as shown in Fig.~2 and the center coordinates
are the constants of motion. 

%
%
\begin{figure}[th]
\centerline{\includegraphics[width=5cm]{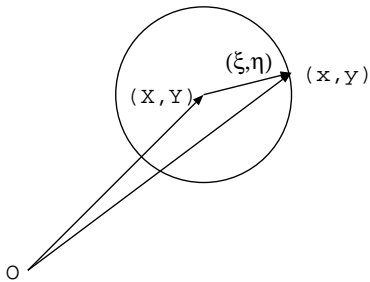}}
\vspace*{8pt}
\caption{A schematic illustration of electron's circular motion around 
the center coordinates $(X,Y)$. The relative coordinates $(\xi,\eta)$ 
are $(x-X,y-Y)$.}
\end{figure}

In quantum mechanics, the center coordinates, $(X,Y)$, become operators 
but satisfy,
\begin{eqnarray}
& &[X,v_x]=[X,v_y]=0,    \nonumber    \\
& &[Y,v_x]=[Y,v_y]=0, 
\end{eqnarray}
hence they commute with the  $H_0$ also. They are the constants of 
motion as in classical mechanics. Since the commutation relation between 
$x$ and $y$, which vanishes, is given as 
\begin{equation}
[x,y]={[v_x,v_y] \over {{\omega}_c}^2}+[X,Y],  
\end{equation}
and the velocity operators do not commute each other due to the magnetic 
field, the center coordinates become also non-commutative. They satisfy 
\begin{eqnarray}
[X,Y]&=&-{[v_x,v_y] \over {\omega_c}^2}\nonumber \\
&=&{i\hbar \over eB}.
\end{eqnarray}
Hence it is impossible to diagonalize both center variables same time. 
They satisfy uncertainty relations. The von Neumann lattice coherent 
state\cite{complete set} is an eigenstate of the operator, $ X+iY $,
 with a complex eigenvalue, $z_{mn}=a(m {\omega}_x+n
{\omega}_y)$, where $a=\sqrt{2\pi\hbar\over eB}$ and $m$ and $n$ are 
integers. A set of coherent states becomes complete under the condition, 
Im$({\omega }_x^*{\omega}_y)=1$. 
They are symmetric in two-directions and has 
minimum uncertainty allowed by the commutation relation. 

The von Neumann lattice coherent state is constructed by $A={{\sqrt \pi}
\over a}(X+iY)$ and its conjugate as, 
\begin{eqnarray}
|\alpha_{m,n}\rangle&=&{\exp}\left(i\pi(m+n+mn)+{\pi}^{1/2}(A^\dagger{z_{mn}
 \over a}-A{z_{mn} \over a })\right)|0 \rangle, \nonumber \\
& &[A,A^\dagger]=1, 
\end{eqnarray}
and satisfy 
\begin{equation}
\langle \alpha_{m+m',n+n'}|{\alpha}_{m',n'}
 \rangle={\exp}\left(i\pi(m+n+mn)-{\pi \over 2}\left|{z_{mn} \over a}
\right|^2\right).
\end{equation}
The above matrix elements of two coherent states have translational 
invariant form. Hence momentum states are defined by Fourier transformation: 
\begin{equation}
\vert{\alpha}_{\bf p}\rangle=\sum_{m,n}e^{ip_xm+ip_yn}
\vert\alpha_{m,n}\rangle,
\end{equation}
\begin{equation}
\langle{\alpha}_{\bf p}\vert{\alpha}_{{\bf p}'}\rangle=
\gamma({\bf p})\sum_{N}(2\pi)^2\delta({\bf p}-{\bf p}'
-2\pi{\bf N}),
\end{equation}
where two components of {\bf N}, $N_x$ and $N_y$, are integer and the 
fundamental region of {\bf p} is defined as, $|p_x| \le\pi, |p_y|\le\pi$. 

The normalization constant is given as, 
\begin{eqnarray}
{\gamma}({\bf p})&=&{\beta}({\bf p})^*{\beta}({\bf p}),
\label{alpha}\\
{\beta}({\bf p})&=&\left(2{\rm Im} {\tau} \right)^{1\over4}
e^{i{ {\tau} \over 4{\pi}}{p_y}^2}
\vartheta_1\left({p_x+{\tau} p_y\over2{\pi}}\left|\right. 
{\tau}\right),
\label{beta}\\
{\tau}&=&-{{\omega}_x \over {\omega}_y}
\end{eqnarray}
where $\vartheta_1(z\vert \tau )$ is a theta function of the first kind, 
and $\beta({\bf p})$ obeys a nontrivial boundary condition
\begin{equation}
\beta({\bf p}+2\pi{\bf N})=e^{i\phi({\bf p,N})}\beta({\bf p}),
\end{equation}
where $\phi({\bf p,N})=\pi(N_x+N_y)-N_y p_x$. 
The norm, $\gamma({\bf p})$, vanishes once in the inside of fundamental 
region, at $\bf p=0$. 

The relative coordinates, $(\xi ,\eta)$, are the coordinates of the electron 
measured from the center, $(X,Y)$, and are proportional to the velocity 
operators,
\begin{equation}
\xi=x-X={v_y \over \omega_c},\quad \eta=y-Y=-{v_x \over \omega_c},
\end{equation}
and commute with the center coordinates, $(X,Y)$. Eigenstates of  $H_0$ are 
those of harmonic oscillator with discrete eigenvalues, $E_l$ of Eq.~(7), 
\begin{equation}
H_0 |f_l \rangle = E_l |f_l \rangle.
\end{equation}
For basis of two dimensional electrons in the magnetic field, we use the 
direct product between eigenstates of $H_0$ of eigenvalue $E_l$, 
$|f_l\rangle$, and the above momentum states, 
\begin{equation}
|l,{\bf p}\rangle=|f_l \rangle \otimes\frac{
|\alpha_{{\bf p}}\rangle}{\beta({\bf p})}.
\end{equation}

\subsection{Field theory}
Using the von Neumann lattice basis we develop a field theory of the 
quantum Hall system. Hamiltonian is given as a summation of the free term, 
${\cal H}_0$ and the interaction term, ${\cal H}_{int}$, 
\begin{eqnarray} 
{\cal H}_{total}&=&{\cal H}_0+{\cal H}_{int}, \\
{\cal H}_0&=&\int d^2 x \psi^\dagger ({\bf x}) H_0 \psi({\bf x}),\\
{\cal H}_{int}&=&\frac{1}{2}\int{{{d^2 k}\over (2 \pi)^2} \rho({\bf k})
V({\bf k})\rho(-{\bf k})},
\end{eqnarray}
where ${\rho{(\bf k)}}$ is the density operator defined as, 
\begin{equation}
\rho({\bf k})=\int d^2x e^{-i \bf k \cdot x}\psi^{\dagger}({\bf x},t) 
\psi({\bf x},t)
\end{equation}
and  $V{(\bf k)}$ is the Coulomb potential, $V({\bf k})=q^2{2\pi \over
k}$, $q^2={e^2 \over 4\pi\epsilon}$, $\epsilon$ is the dielectric constant
of matter. The disorder term, 
${\cal H}_{disorder}$, is added later.

Action of the system is given as, 
\begin{equation}
S=\int dt d^2 x [\psi^{\dagger}({\bf x},t) i \hbar 
{\partial \over \partial t}\psi({\bf x},t)]-{\cal H}_{total},
\end{equation}
and is written as, 
\begin{equation}
S=\int dt \int_{\rm BZ} {d^2 p \over (2 \pi)^2}[b_l^{\dagger}({\bf p},t) 
i \hbar {\partial \over \partial t} b_l({\bf p},t) ]-{\cal H}_{total},
\end{equation} 
when electron field is expanded with the above basis as, 
\begin{equation}
\psi({\bf x},t)=\int_{{\rm BZ}}{d^2p\over(2\pi)^2}\sum_{l=0}
^{\infty}b_l({\bf p},t)\langle {\bf x}|l,{\bf p}\rangle,
\end{equation}
where BZ in the integration region in momentum stands for the integration in 
the fundamental region. For simplicity, we omit $t$ in the operators 
unless it is confusing hereafter. Expansion coefficients, $b_l({\bf p})$, 
and their conjugates, $b^\dagger_{l}({\bf p})$, 
are the annihilation and creation operators of electrons in Landau levels 
which satisfy equal time anti-commutation relations, 
\begin{equation}
\{b_{l}({\bf p}),b^\dagger_{l'}({\bf p}')\}=\delta_{l,l'}
\sum_N (2\pi)^2\delta({\bf p}-{\bf p}'-2\pi{\bf N})e^{
i\phi({\bf p',N})}.
\end{equation}
Operators $b^\dagger_l({\bf p})$ and $b_l({\bf p})$ operate on the 
many-body states and satisfy a torus boundary condition with a phase factor, 
$\phi({\bf p,N})=\pi(N_x+N_y)-N_y p_x$, in the momentum space. 

The free many-body Hamiltonian becomes diagonal and is given as, 
\begin{equation}
{\cal H}_0=\sum_{l=0}^{\infty}\int_{{\rm BZ}}{d^2 p\over (2\pi)^2}E_l
 b^\dagger_{l}({\bf p})b_{l}({\bf p}).
\end{equation}
Due to degeneracy of Landau levels, one-particle energy does not depend on 
its momentum. The density operator is expressed as, 
\begin{eqnarray}
\rho({\bf k}) &=&\sum_{l,l'=0}^{\infty} \int_{{\rm BZ}}{d^2 p\over (2\pi)^2}
b^\dagger_{l}({\bf p}) b_{l'}({\bf p}+a{\bf \hat k})f_{l l'}({\bf k})
\nonumber \\ 
& &\times \exp[{i \over 4 \pi} a \hat k_x(2 p_y+a \hat k_y)],\\
{\bf j} ({\bf k}) 
&=&
\sum_{l,l'=0}^{\infty} \int_{\rm BZ} {d^2p\over (2\pi)^2}
 b^\dagger_{l} ({\bf p}) b_{l'} ( {\bf p} + a{\hat {\bf k}} )
\langle f_l| {1 \over 2} \{ {\bf v},e^{-i(k_x \xi+k_y \eta)}\}
|f_{l'} \rangle \nonumber \\ 
& &\times \exp[{i \over 4 \pi} a \hat k_x(2 p_y+a \hat k_y)],   \\
f_{l l'}({\bf k})&=&\langle f_l|e^{-i(k_x \xi+k_y \eta)}|f_{l'} \rangle 
\end{eqnarray}
here, ${\hat k}_i=W_{ij}k_j $ with the matrix defined by
\begin{equation}
W=\left(
\begin{array}{cc}
{\rm Re}[\omega_x]  &  {\rm Im}[\omega_x] \\
{\rm Re}[\omega_y]  &  {\rm Im}[\omega_y] \\
\end{array}
\right).
\end{equation}
For rectangular  lattice $\omega_x=r_s$, $\omega_y={i \over r_s}$, where
$r_s$ is an asymmetry parameter. 
It is convenient to introduce a unitary matrix,
\begin{equation}
U^\dagger_{ll'}({\bf p})=\langle f_l\vert
e^{i({\tilde p_x \xi+\tilde p_y \eta})/a-{i\over4\pi}p_xp_y}
\vert f_{l'}\rangle,
\end{equation}
here $\tilde p_i=W^{-1}_{ij}p_j$,
and to transform electron operators as,
\begin{equation}
\tilde b_l({\bf p})=\sum_{l'}U_{ll'}({\bf p})b_l({\bf p}).
\end{equation}
The density operator is written in the diagonal form and 
the current operator are written also in simple forms with the transformed 
operators,
\begin{eqnarray}
\rho({\bf k})&=&\int_{{\rm BZ}}{d^2 p\over (2\pi)^2}\sum_
{l}\tilde b^\dagger_{l}({\bf p})\tilde b_{l}({\bf p}+a\hat{\bf k}),
\nonumber \\
{\bf j}({\bf k})&=&\int_{{\rm BZ}}{d^2 p\over (2\pi)^2}
\sum_{l,l'}
{\tilde b}_{l}^\dagger({\bf p})\langle f_l|{{\bf v}+{a\omega_c\over2\pi}(
\tilde{\bf p}+{a\over2}{\bf k})}| f_{l'} \rangle 
\tilde b_{l'}({\bf p}+a\hat{\bf k}),
\end{eqnarray}
where ${\bf v}$ is the velocity operator and its components are proportional 
to the relative coordinates. Commutation relation of charge density with the 
electron operators is equivalent to that of local field theory when the 
transformed operators are used and is given as, 
\begin{equation}
 [\rho({\bf k}),\tilde b_{l}({\bf p})] =-\tilde b_{l}({\bf p}+a{\hat{\bf k}}),
\end{equation}
if the momentum $\bf k$ and $\bf p$ are in the fundamental region. 
From these relations, Ward-Takahashi identity\cite{ward-takahashi} 
between the vertex part and the propagator is derived.

Let us define the one-particle irreducible vertex part, $\tilde{\Gamma}^\mu$ 
from time ordered product as, 
\begin{eqnarray}
\int dz_0 dx_0 dx'_0 e^{iq_0z_0-ip_0x_0+ip'_0x'_0}\langle 
T( j^\mu (z_0,{\bf q})\tilde{b}_l(x_0,{\bf p})\tilde{b}_{l'}^\dagger 
(x'_0,{\bf p'}))\rangle \nonumber\\
= (2\pi)^3 \delta(p+Q-p')
\tilde{S}_{ll_1}(p)\tilde{\Gamma}_{l_1l_2}^\mu (p,p+Q)
\tilde{S}_{l_2l'}(p+Q),
\end{eqnarray}
where $j_{\mu}(x)=({\rho}(x),{\bf j}(x))$ and $Q^{\mu}=
(q_0,a\hat{q}_x,a\hat{q}_y)={t^\mu}_\nu q^\nu$ is a linear combination of 
$q^\mu$ and $\tilde{S}$ is the full propagator defined by 
\begin{equation}
\int dx dx'e^{ipx-ip'x'}\langle T(\tilde{b}_l(x)\tilde{b}_{l'}^\dagger (x'))
\rangle = (2\pi)^3 \delta(p-p')\tilde{S}_{ll'}(p).
\end{equation}
From the current conservation and the commutation relation, 
the Ward-Takahashi identity\cite{ward-takahashi} between $\tilde{S}$ and 
$\tilde{\Gamma}^\mu$, 
\begin{equation}
\tilde{\Gamma}_{\mu}(p,p)
={t^\nu}_\mu {\partial \tilde{S}^{-1}(p) \over \partial p^{\nu}},
\label{wt}
\end{equation}
is satisfied. Using the current correlation function,
\begin{equation}
\pi^{\mu\nu}(q,q')=\int dx dx'e^{iqx-iq'x'}\langle T(j^\mu (x)  j^\nu (x'))
\rangle=(2\pi)^3 \delta(q-q')\pi^{\mu\nu}(q)+\pi^{\mu\nu}_{(2)}(q,q'), 
\end{equation}
the Hall conductance is defined by the slope of the momentum conserving part, 
$\pi^{\mu\nu}(q)$, at the origin and is written as 
\begin{equation}
\sigma_{xy}={e^2 \over 3!}\epsilon^{\mu\nu\rho}
\partial_{\rho}\pi_{\mu\nu}(q)|_{q=0},
\label{hallc}
\end{equation}
where $\epsilon_{\mu\nu\rho}$ is anti-symmetric tensor. As we will see later 
the momentum non-conserving term, $\pi^{\mu\nu}_{(2)}(q,q')$, does not 
contribute to the conductance. 

Using the above relations we are able to write the Hall conductance as a 
particular form of a topological invariant\cite{topological} 
defined, 
\begin{equation}
\sigma_{xy}={e^2 \over h} N_w,
\end{equation}
\begin{equation}
N_w={1\over 24\pi^2}\int_{{\rm BZ}\times S^1}
d^3 p \epsilon_{\mu\nu\rho} {\rm tr} \left(
\partial_\mu \tilde{S}^{-1}(p)\tilde{S}(p)
\partial_\nu \tilde{S}^{-1}(p)\tilde{S}(p)
\partial_\rho \tilde{S}^{-1}(p)\tilde{S}(p)\right)
\label{topological}.
\end{equation}
In the above equation, the integration region in the spatial 
components is a torus defined from BZ and  the integration region in the 
energy variable is equivalent to $S^1$ when the propagator has 
imaginary part that is defined from the Fermi energy. The final 
formula is valid in almost arbitrary quantum Hall systems of 
interactions and disorders. 

It is rather trivial to derive the final expression Eq.~(\ref{topological}) 
in the free system. In the free system, the slope of the current correlation 
function at the origin is given by the one loop diagram with three vertecies 
with zero external momenta, as shown in Fig.~3. Using the Ward-Takahashi 
identity, the vertex parts are replaced with the derivatives of the inverse 
of the propagator, Eq.~(\ref{wt}). So we end up with the final formula, 
Eq.~(\ref{topological}). In general systems that have interactions and 
disorders, it is not so trivial to derive this formula that 
further arguments will be given. 
%
%
\begin{figure}[th]
\centerline{\includegraphics[width=4cm]{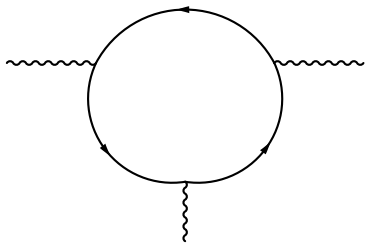}}
\vspace*{6pt}
\caption{Feynman diagram of three external current lines with zero momentum 
in the lowest order. The solid lines correspond to the electron propagator 
and the wavy lines correspond to the electromagnetic current.}
\end{figure}

$N_w$ is a winding number of the mapping from momentum space to the space of 
matrix defined by the propagator, $\tilde{S}(p)$, and agrees with an integer 
if the integrand is single-valued and has no singularity in the integration 
region. This condition is satisfied in the gap region where the ground state 
is isolated in the energy and excited states have  energy gaps and in the 
localized state region where one-particle states have localized wave 
functions and discrete energies as well.  

%
%
\begin{figure}[th]
\centerline{\includegraphics[width=4cm]{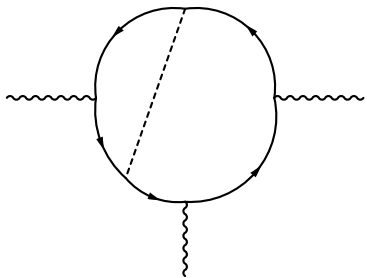}}
\vspace*{6pt}
\caption{Feynman diagram of three external current lines with zero momentum 
in the second order correction. The solid lines correspond to the electron 
propagator and the wavy lines correspond to the electromagnetic current. 
The dashed line shows the Coulomb interaction and one of the vertex parts 
is modified.}
\end{figure}

\subsection {Interactions}

In the systems of interactions, it is highly non-trivial to see the 
fact that the Hall conductance is written by the topological winding 
number of the propagator. We give a proof of this fact by treating the 
interaction term perturbatively. The electron propagator is defined by 
${\cal H}_{0}$ and corrections are caused by interaction Hamiltonian. 
To use diagrammatic analysis, we assign a solid line for the electron 
and a dashed line for a longitudinal photon, which expresses 
Coulomb interaction. 

The current correlation function, $\pi_{\mu \nu}(q,q')$, is expressed with 
electron propagator and vertex parts defined by the current operators and 
others. It was shown in the previous subsection that the slope of lowest 
order amplitude was obtained by the above topological formula, if 
Ward-Takahashi identity is used. Corrections are calculated by using 
perturbative expansions with respect to ${\cal H}_{int}$. The first order 
correction to the current correlation function and its slope at the origin 
of the momentum  are given in Feynman diagrams of Fig.~4. Thus the total 
amplitude is written by using the propagator and the vertex part. 
By the Ward-Takahashi identity, they are connected each other and the 
slope of the total amplitude is written by the topological winding number 
of the full propagator.
%
%
\begin{figure}[th]
\centerline{\includegraphics[width=4cm]{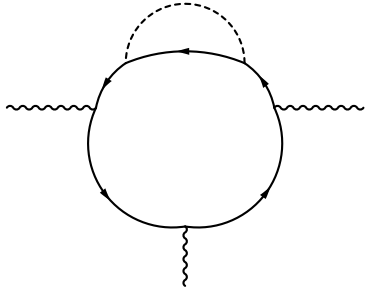}}
\vspace*{6pt}
\caption{Feynman diagram of three external current lines with zero momentum 
in the second order correction. 
The solid  lines correspond to the electron propagator and the wavy
lines correspond to the electromagnetic current. 
The dashed line shows the Coulomb interaction and 
one of the electron propagators is modified.}
\end{figure}

Similarly the slope of the current correlation function at the origin in 
higher orders is classified to diagrams of two different 
topologies.\cite{von neumann2} In first type of diagrams, 
three vertices are attached to three different  electron lines, as is given 
in Fig.~6. Using the Ward-Takahashi identity, integrand of these amplitudes 
can be written as the third derivative with respect to all three components 
of internal electron momentum. These amplitudes are symmetric in the 
Lorenz indecis ${\mu}$, ${\nu}$, and ${\rho}$ of the currents and do not 
contribute to the asymmetric part in ${\mu}$, ${\nu}$, and ${\rho}$. 
Since the Hall conductance is the asymmetric part, the Hall conductance 
is not affected at all from this kind of amplitude. 

%
%
\begin{figure}[th]
\centerline{\includegraphics[width=4cm]{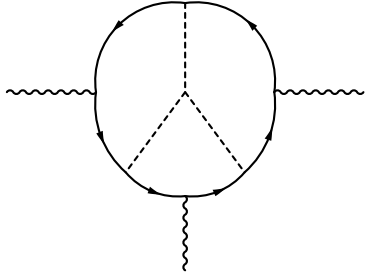}}
\vspace*{6pt}
\caption{First type of Feynman diagram of three external current lines 
with zero momentum in higher order corrections. The solid lines correspond 
to the electron propagator and the wavy lines correspond to 
the electromagnetic current. The dashed line shows an electron interaction 
which is connected with all electron propagators.}
\end{figure}

In second type of diagrams, three external vertices are attached to one 
electron line which include all higher order effects, as is given in 
Fig.~7. So this amplitude is written by the full propagator and full 
vertices. Since the current conservation and the equal time commutation 
relation between the charge density and electron operators are satisfied, 
the Ward-Takahashi identity between the full vertex part and the full 
propagator is satisfied as well. Hence the slope of the amplitude of the 
second class is written by the topological invariant 
Eq.~(\ref{topological}) of the full propagator. 

%
%
\begin{figure}[th]
\centerline{\includegraphics[width=4cm]{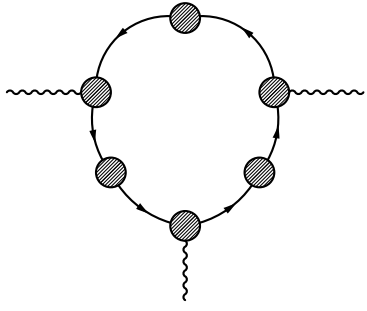}}
\vspace*{6pt}
\caption{Second type of Feynman diagram of three external current lines 
with zero momentum in higher order corrections. 
The solid lines correspond to the electron propagator and the wavy lines 
correspond to the electromagnetic current. The dashed lines correspond 
to interactions.The propagators and the vertices are modified by 
interactions and satisfy extended Ward-Takahashi identity.}
\end{figure}

It is possible to extend this theorem to general systems which have other 
dynamical freedoms than the electrons such as phonon and others. Namely 
the Hall conductance is written by the same topological expression in quite 
general systems. For later convenience, we study the current correlation 
function with two different momenta in the currents, as is given in 
Fig.~8.  

%
%
\begin{figure}[th]
\centerline{\includegraphics[width=4cm]{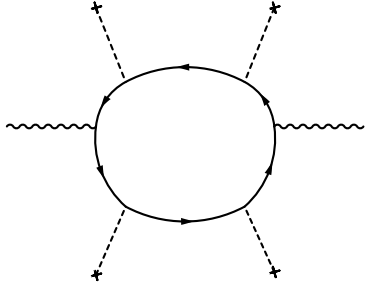}}
\vspace*{6pt}
\caption{Feynman diagram of momentum non-conserving current correlation 
function with small different momenta of the currents. 
The solid lines correspond to the electron propagator and the wavy lines 
correspond to the electromagnetic current. The dashed lines correspond to 
dynamical freedoms which are carrying momentum and the momentum are not 
conserved. }
\end{figure}

Let us define such current correlation function, 
$\pi^{\mu\nu}_{(2)}(q,q')$, which is a function of two momenta $q_{\mu}$ and 
$q'_{\mu}$ and satisfies the identities,
\begin{equation}
q_{\mu}\pi^{\mu\nu}_{(2)}(q,q')=\pi^{\mu\nu}_{(2)}(q,q')q'_{\nu}=0.
\label{momentum-nonconserving1}
\end{equation}  
Since the two momenta are independent, it is possible to differentiate these 
equations with respect to $q^{\rho}$ or $q'^{\rho}$. Then we have, 
\begin{eqnarray}
\pi^{\rho\nu}_{(2)}(q,q')+  q_{\mu} \partial_{\rho}
 \pi^{\mu\nu}_{(2)}(q,q')
&=&0,
\nonumber \\
\pi^{\mu\rho}_{(2)}(q,q')+  q'_{\nu} \partial'_{\rho}
 \pi^{\mu\nu}_{(2)}(q,q')
&=&0.
\label{momentum-nonconserving2}
\end{eqnarray}   
Thus if the amplitude is smooth around the origin in two momenta and the 
first derivatives are finite, the amplitude is proportional to two momenta, 
$q^{\mu}$, and $q'^{\nu}$ \cite{coleman-hill}. Thus these amplitudes do 
not contribute to the linear slope at the origin. If a momentum conserving 
amplitude is written as a limit of momentum non-conserving amplitude, this 
amplitude also does not contribute to the linear slope.  

In situations where the ultraviolet divergences are involved 
in perturbative expansions, the renormalized propagator and vertex are to be 
used. They are defined in such manner that satisfy the Ward-Takahashi 
identity, and thus the Hall conductance is written by the topological 
expression of the renormalized propagator. 

Thus the quantized Hall conductance is not affected by 
interactions.\cite{topological,topological-2} 

\subsection{ Random disorders} 

In systems of disorders, translational invariance is 
lost and momentum becomes not a good quantum number. It seems that the 
momentum conserving term does not exist in the system of disorders. 
However we will see that it is not the case and the momentum conserving 
term exists. To study disorder effects, we use the diagrammatic analysis. 
It is shown that the above topological expression is valid also. For the bulk 
quantity such as conductance which is defined by a ratio between total 
voltage and total current, momentum conserving term contributes. 
Momentum becomes a good quantum number even in the systems of disorders, 
as far as the bulk physical quantities are concerned. 

If the momentum is not conserved, the current correlation function, 
$\pi^{\mu\nu}(q,q')$, becomes a function of two momenta and satisfies the 
identities, Eq. (\ref{momentum-nonconserving1}), and 
Eq. (\ref{momentum-nonconserving2}). As was seen in the previous part, 
if the amplitude is smooth around the origin in two momenta and the 
first derivatives are finite, the amplitude is proportional to two momenta, 
$q^{\mu}$ and $q'^{\nu}$. The condition of the smoothness around the origin 
of the momenta is equivalent to the absence of infrared divergence and is 
satisfied in the energy gap region and in the localized state region since 
one-particle states have the energy gap or the one-particle wave functions 
have finite spatial extensions.


Disorder potentials is expressed in the Hamiltonian: 
\begin{equation}
{\cal H}_{dis}
=\int{{d^2 k} \rho({\bf k})V_{dis}({\bf k})}.
\end{equation}
Spectrum of ${\cal H}_0+{\cal H}_{dis}$ becomes different from that of
$H_0$ and depends 
on the potential $V_{dis}$. For random potentials, energy eigenstates  
appear in the energy regions between the Landau levels. From numerical 
study it is known that these states have discrete energy and localized wave 
functions. So they do not contribute to conductance. Hence when the Fermi 
energy is located in this region, we treat random potential ${\cal H}_{dis}$ 
perturbatively in the calculation of conductance. For the case of periodic 
potentials, perturbative treatment is not good and nonperturbative treatments 
for the calculations of the energy spectrum and conductance are to be made. 

Now we give a proof that the momentum non-conserving term does not 
contribute to the conductance. Systems  with   the random potentials are 
studied in the following. Expectation value of the current density is 
connected with the current correlation function as, 
\begin{equation}
j_{i}({\bf x})=\int {d^2 x'} \pi_{i0}({\bf x},{\bf x'})A_{0}({\bf x'}).
\end{equation}
The total current is obtained by  integrating the current density in whole 
spatial region,
\begin{equation}
I_{i}=\int {d^2 x} {j_{i}({\bf x})}.
\end{equation}
Let us substitute the Fourier transformation of the current correlation 
function and the scalar potential for the coordinate space to the momentum 
space. Then we have, 
\begin{equation}
I_{i}= {\partial \over \partial q'_{\rho}} \pi_{i0}({\bf q},{\bf q'})
|_{q=q'=0}V,
\end{equation}
where $\rho$ is different from $i$ and V is the potential difference
expressed with  $A_0$. Consequently the coefficient ${I_{i} \over V}$ is 
proportional to the linear term in the momentum. Because the momentum 
non-conserving term has higher power than linear, the non-conserving term 
does not contribute to the conductance. 

Now ${\cal H}_{dis}$ induces a finite flow of momentum and external momenta 
carried by the two currents  is not conserved generally. This amplitude 
depends on two independent momenta and behaves as bilinear in the momenta. 
Hence the momentum non-conserving term does not contribute to the 
conductance. The momentum conserving terms are generated by cancellation 
of the momentum flow  due to disorders and contribute to the conductance.  
These momentum conserving terms are similar to those of the system of 
interactions in which dynamical freedoms carry the momentum and the momenta 
of the currents disagree. Hence by introducing  new interaction terms 
corresponding to each sets of momentum conserving diagrams, the previous 
arguments of applying current conservation, commutation relation, and 
Ward-Takahashi identity are applicable also in the systems of disorders.
    
\subsection{Periodic potential}
 
If the potential is periodic in space, one-particle states form band 
structures\cite{hofstadter,duality-periodic}. Since magnetic field gives
a periodic structure by itself, which can be seen with von Neumann lattice 
representation clearly, band structure is very sensitive to the ratio of 
the potential period with the magnetic length. Peculiar structure was 
identified in one particle spectrum. Hall conductance behaves also in a 
peculiar way\cite{tknd}.     

Generally, a periodic potential lifts the degeneracy of the 
Landau level. In what follows, we assume that the potential lattice  is 
formed by two linear-independent basis vectors with integer coefficients. 
The periodic potential with this property is written as\cite{von neumann2} 
\begin{equation}
V({\bf x})=\sum_{{\bf N}\in {\bf Z}}
v({\bf x}+aN_x {\bf w}_x^{pot}+aN_y  {\bf w}_y^{pot}),
\end{equation}
where ${\bf w}_x^{pot}$, ${\bf w}_y^{pot}$ are the basis vectors of 
the potential lattice with the components 
(Re$ \omega_x^{pot}$, Im$\omega_x^{pot}$) and 
(Re$\omega_y^{pot}$, Im$\omega_y^{pot}$). 
As is well-known, a periodic potential problem in a magnetic field is 
very sensitive to the flux penetrates a unit cell of the potential lattice. 
The unit cell means a parallelogram spanned by basis vectors of the 
potential lattice. If the flux $\Phi$ is a rational multiple 
of the flux quantum $\Phi_0={h \over e}$, 
\begin{eqnarray}
{\Phi\over \Phi_0}&=&{\rm Im}[(\omega_y^{pot})^*\omega_x^{pot}] \nonumber \\
&=&{q\over p}, 
\label{ratio}
\end{eqnarray}
each  Landau band splits into $q$ sub-bands. 
The fundamental region of the momentum is reduced to one $q$-th. The 
spectrum in the reduced region of the momentum is $p$-fold degenerate. 

When the flux is given by Eq.~(\ref{ratio}), it is convenient  to select  
the basis vectors of the von Neumann lattice as $\omega_x=\omega_x^{pot}/q, 
\omega_y=p\omega_y^{pot}$. The moduli of the von Neumann lattice becomes 
$\tau =\tau^{pot}/pq$, where the moduli of the potential lattice is defined 
by $\tau^{pot}=-\omega_y^{pot}/\omega_x^{pot}$. In this case, 
the potential energy term in the second quantized form becomes
\begin{eqnarray}
H_{\rm pot}&=&{1\over q}\sum_{l,l'}\sum_{s=0}^{p-1}\sum_{r=0}^{q-1}
 \int d^2 x v({\bf x})
\int_{{\rm RBZ}}{d^2 p \over (2\pi)^2}
b^\dagger_{l}({\bf p})b_{l'}(p_x-2\pi{r\over q},p_y) \nonumber \\
  & &\times \langle l,p_x-2\pi \left({\tilde{y}\over a}+{s\over p}\right),
              p_y+2\pi {\tilde{x}\over a}|{\bf x}={\bf 0}\rangle \nonumber \\
        & & \times \langle {\bf x}={\bf 0}|l',p_x-2\pi\left({\tilde{y}\over a}+
                         {s\over p}+{r\over q}\right),
                         p_y+2\pi {\tilde{x}\over a} \rangle,\\
\tilde x_i=x_j W^{-1}_{ji}, \nonumber
\end{eqnarray}
where RBZ is the reduced region of the momentum, 
$|p_x|\le \pi/q$, $|p_y|\le \pi$. The eigenvalue equation reads 
\begin{eqnarray}
& &{1\over q} \sum_{l'}\sum_{s=0}^{p-1}\sum_{r'=0}^{q-1}  \int d^2x\;\; 
v({\bf x})\langle l,p_x-2\pi\left({\tilde{y}\over a}+{s\over p}+
{r\over q}\right),
                    p_y+2\pi {\tilde{x}\over a}|{\bf x= 0}\rangle
                   \nonumber \\
& & \times  \langle {\bf x=0}|l',p_x-2\pi\left({\tilde{y}\over a}+
                         {s\over p}+{r' \over q}\right),
                         p_y+2\pi {\tilde{x}\over a} \rangle
\psi_{l'r'} ({\bf p})=(E-E_l)\psi_{lr}({\bf p}) ,
\label{eigen}
\end{eqnarray}
The function $\psi_{lr}({\bf p})$ is given by the following form: 
\begin{equation}
\psi_{lr}({\bf p})=\psi_l (p_x-2\pi{r\over q},p_y).
\end{equation}
The equation is solved by diagonalizing a $Lq\times Lq$ matrix, where 
$L$ is the number of Landau levels. Therefore, each Landau band splits 
into $q$ sub-bands generally. 
It is easy to see that the spectrum of the eigenvalue equation
is invariant under the  translations  $p_x \to p_x+2\pi n/p$ 
and $p_x \to p_x+2\pi n/q$, where $n$, $m$ are integers. 
Since $p$, $q$ are coprime integers, the above symmetry reads 
\begin{equation}
E({\bf p})=E(p_x+2\pi{n\over pq},p_y).
\end{equation}
Thus, it is proven that the spectrum in RBZ are $p$-fold degenerate.

Next, we consider a system with defects and a periodic potential. 
To extract properties in such a system, let us suppose a defect 
in a periodic short-range potential. 
That is, the potential is given by 
\begin{equation}
V({\bf x})=\sum_{{\bf N}\in {\bf Z}}
V_0 a^2 \delta ({\bf x}-aN_x {\bf w}_x^{pot}-aN_y  {\bf w}_y^{pot}) 
+g a^2 \delta ({\bf x}-aM_x {\bf w}_x^{pot}-aM_y  {\bf w}_y^{pot}), 
\end{equation}
for  integers  $M_x$, $M_y$. The second term breaks the periodicity of the 
potential. If we neglect the second term, the eigenvalue equation becomes 
\begin{equation}
\sum_{l'r'}{V_0\over q}(D_l^\dagger ({\bf p})D_{l'}({\bf p}))_{rr'}
\psi_{l'r'}({\bf p})=(E-E_l)\psi_{lr}({\bf p}),
\label{antidot}
\end{equation}
where $p\times q$ matrix $D_l$ is given by 
\begin{equation}
(D_l({\bf p}) )_{sr}=a\langle {\bf x=0}|l,p_x-2\pi \left(
{s\over p}+{r\over q}\right),p_y 
\rangle  \qquad (s=0,..,p-1; r=0,..,q-1).
\end{equation}
The spectrum in the LLL consists of flat bands and Hofstadter-type bands. 
The wave function of the flat band at $E=E_l$ is given by 
$\psi_k=\delta_{kl} Ker(D_l)$. 

In Fig.~9 the spectrum for the square lattice potential 
is shown. We observe that self-similar pattern and large gap above the flat 
bands exist in this figure. Hofstadter-type  bands  above flat bands tend 
to a set of bound states as $t$ becomes infinity. 
%
%
\begin{figure}[th]
\centerline{\includegraphics[width=11cm]{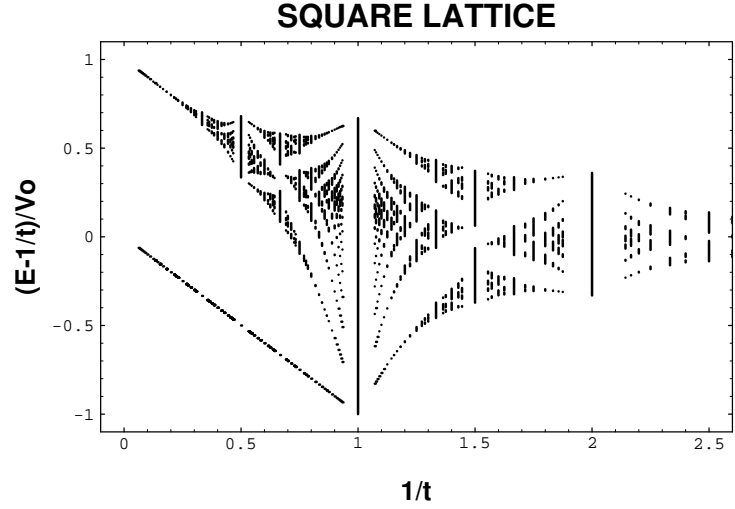}}
\vspace*{8pt}
\caption{One-particle spectrum of the periodic array of short range potentials
that resembles Hofstadter butterfly. $t$ is ${\Phi_0 \over \Phi}$ and 
on-site energy is subtracted from the one-particle energy. 
}
\end{figure}

Incorporating the defect leads to the following additional term in the L.H.S. 
of Eq.~(\ref{antidot}): 
\begin{equation} 
g \sum_{l'm'}\int_{{\rm RBZ}} 
{d^2 q\over (2\pi)^2} (D_l^\dagger ({\bf p}))_{rs} 
( D_{l'} ({\bf q}))_{sm'} e^{iqM_x(q_x-p_x)+iS(q_y-p_y)}\psi_{l'm'} ({\bf q}),
\end{equation}
where ${\rm Mod}(M_y,p)=s$ and $(M_y-s)/p=S$. 
Apparently, flat bands is still flat even if the defect exists. 
Furthermore, in each  sub-band gap of the periodic short-range potential 
problem a bound state appears rearranging  eigenstates of  
Hofstadter-type bands. 
The equation of the bound state energies is given by 
\begin{equation} 
g\sum_A \int_{{\rm RBZ}} {d^2 p\over (2\pi)^2} 
{|\sum_{lr} (D_l ({\bf p}))_{sr} {\psi_{lr}}^A({\bf p})e^{iqM_xp_x+iSp_y}|^2 
 \over E-E_A ({\bf p})}=1,
\end{equation}
where  ${\psi_{lr}}^A ({\bf p})$ and $E_A ({\bf p})$ are 
the eigenfunction and the eigenvalue of Eq.~(\ref{antidot}). 
The  R.H.S of the above equation generally becomes infinite 
when $E$ approaches to  the upper edge of the Landau sub-band. 
Also, it becomes minus infinite when $E$ approaches to the lower edge.
Thus, there exist a solution  in each sub-band gap. 
The solution corresponds to a bound state trapped at the defect. 
If many defects exist in the periodic potential, many bound states appear 
in each the  sub-band gap. As the number of defects  increases, 
the sub-band gaps tend to be  filled with the bound states and 
the number of the extended states decreases. 

\subsection{Value of topological winding number and integer quantum Hall 
effect}

The winding number of the propagator is a topological invariant and is stable 
under small changes of the systems. Since the Hall conductance is 
proportional to the winding number that depends upon the propagator and 
the position  of the Fermi energy, the value is computed from Eq.~(50) 
and Eq.~(51) and is given in Fig.~10 as a function of the the Fermi 
energy\cite{von neumann1}. The value stays at  an integer in finite energy 
region and has step-like behaver as a whole. This occurs because one-particle 
energy takes discrete values, $E_l$, and there is no state between these 
energy values in free theory. Hence the plateau in the $E_F$ does not mean 
the plateau in the density. 
%
%
\begin{figure}[th]
\centerline{\includegraphics[width=7cm]{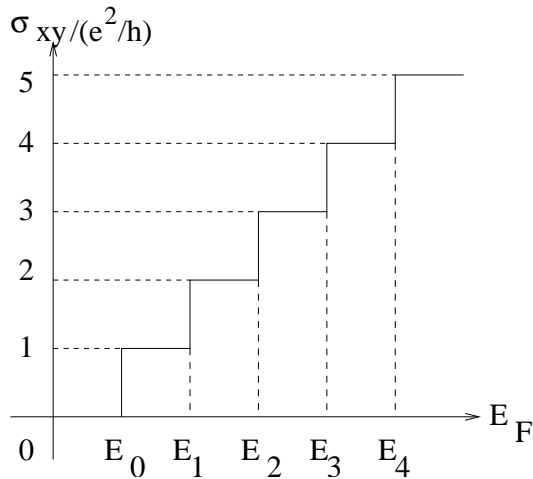}}
\vspace*{8pt}
\caption{The  value of the Hall conductance  of the free electron 
in the magnetic field as a function of the Fermi energy.}
\end{figure}

The situation becomes different and the plateau with respect to the 
density appears in the system of disorders because the topological 
invariant is stable and is unchanged under small changes of the system. 
In systems with disorders, new one-particle states appear in the energy 
region between Landau level's energies. They are localized 
states\cite{localization} and do not contribute to the topological invariant. 
By applying the previous arguments we see that the value agrees exactly with 
that of free theory if the Fermi energy is in the localized state region. 
Although these localized states do not contribute to the value of 
topological invariant, they contribute to the total electron number. 
Consequently the topological invariant stays at one integer as total 
electron number is varied. Thus the Hall conductance has plateaus as a 
function of the electron density in the localized state region as is given 
in Fig.~(11). By treating  interactions perturbatively, we see that this 
is applied also to the system of interactions as far as the ground state 
is in the same phase. 
%
%
\begin{figure}[th]
\centerline{\includegraphics[width=7cm]{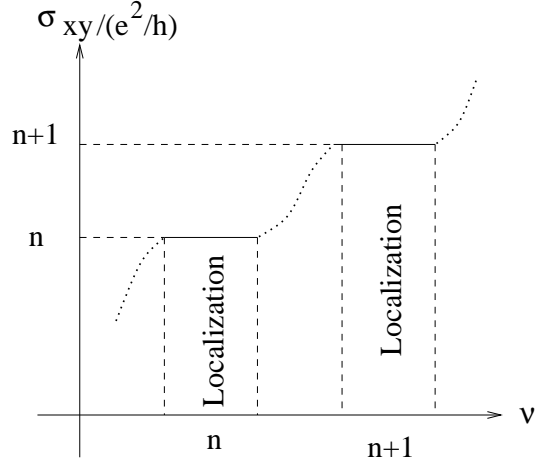}}
\vspace*{8pt}
\caption{The value of Hall conductance of the electrons
 in the disorder potentials  in addition to the magnetic field as a function 
of the filling factor. Plateau appears in localized states region as a 
function of electron density.}
\end{figure}
     
The propagator is modified completely if the periodic potentials is in the 
system. The value of the topological invariant becomes also very different 
from that of the free systems. Our method based on von Neumann lattice 
representation is useful for its calculations. We refer to our 
work\cite{von neumann2} for more details. It is possible to derive Chern 
number formula of  Thouless et. al.\cite{tknd} for the periodic system 
by starting from our formula in  periodic systems. Advantage of our formula 
is that our formula is applicable to the systems with interactions and 
disorders. Fig.~12 gives the Hall conductance in the periodic system. 
%
%
\begin{figure}[th]
\centerline{\includegraphics[width=7cm]{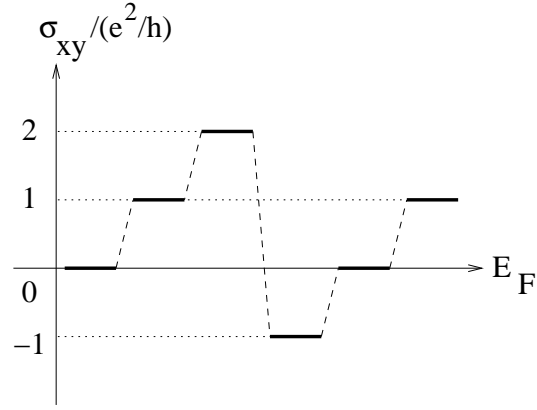}}
\vspace*{8pt}
\caption{The value of Hall conductance of the electrons 
in the periodic potentials as a function of the Fermi energy for $t=5/4$. 
Plateau appears and the value becomes integer multiple of 
${e^2 \over h}$ in a fractional filling factor. }
\end{figure}
      
In the system of Hall gas where one particle energy is not degenerate but
has a dependence upon its momentum,the Hall conductance is unquantized. The
value changes monotonically with the electron density as is given in Fig.~13. 
%
%
\begin{figure}[th]
\centerline{\includegraphics[width=7cm]{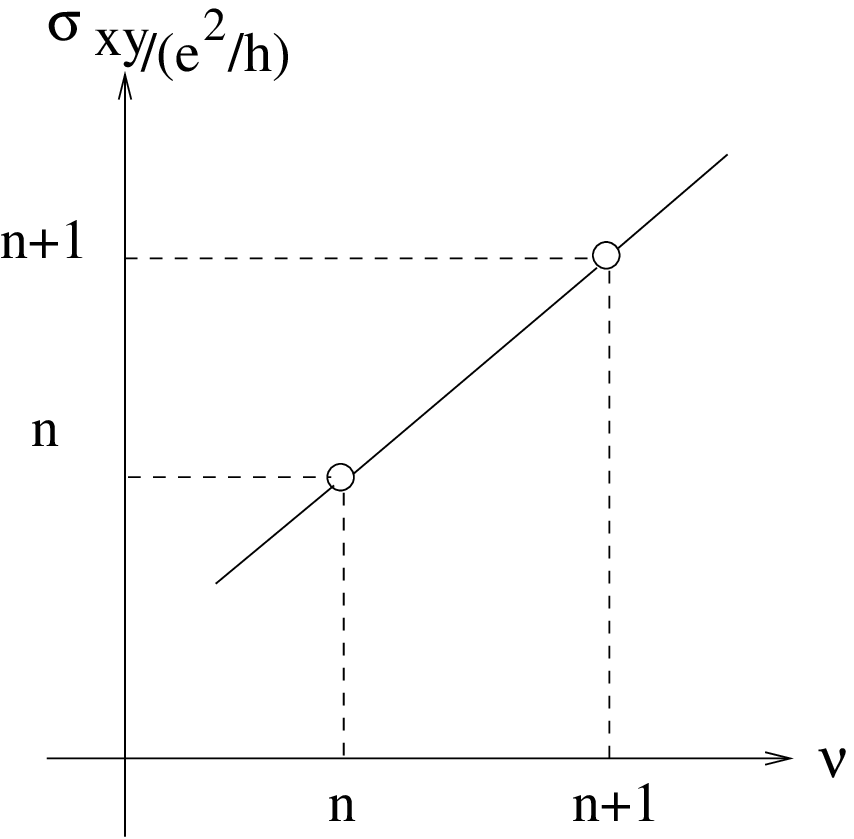}}
\vspace*{8pt}
\caption{The value of Hall conductance  of the electrons in the 
compressible Hall gas region as a function of the filling factor. 
Plateau disappears and the value changes uniformly with the filling factor. }
\end{figure}
      
\section{Anisotropic Hall gas (stripe) state}  

In this section we study an exotic compressible Hall state, anisotropic Hall 
gas state which is compressible and has huge anisotropic resistances. 
We apply mean field theory in von Neumann lattice 
representation\cite{stripe-hf1} which is a convenient representation for 
studying Hall gas state, since the lattice translational symmetry is 
preserved in compressible state. Charge density of anisotropic Hall gas, 
however, will be found to be unidirectional i.e., uniform in one direction 
and periodic in other direction. So the state is equivalent to 
unidirectional charge density wave obtained by Koulakov, Fogler, and 
Shklovskii\cite{stripe1} and Moessner and Chalker\cite{stripe2} at around 
half-filling of higher Landau levels. The state is called stripe state 
sometimes. Recently there are investigations on  several phases such as 
unidirectional charge density wave\cite{stripe-hf2}, asymmetric charge 
density wave \cite{stripe-hf3}, and liquid crystals\cite{stripe-hfx}. 
Numerical methods on small systems have been applied 
also\cite{stripe-numerical}.  We concentrate on the anisotropic 
quantum Hall gas in this article.
We set $\hbar=1$ in this section for simplicity. 

\subsection{Algebraic property}

We study the quantum Hall system with the interaction described by the 
Hamiltonian,  
\begin{equation}
{\cal H}={\cal H}_0+{\cal H}_{int}.
\end{equation}
In this system, the generators of translation in the $y$-direction, 
translation in the x-direction, rotation, and total charge are 
given as,
\begin{eqnarray}
&Q_X&=r_s\sum_{l}  \int_{{\rm BZ}}{d^2p\over (2\pi)^2} b^\dagger_{l}
({\bf p})\left( i {\partial 
\over \partial p_x}-{p_y \over {2 \pi}}\right) b_{l}({\bf p}),\nonumber \\
&Q_Y&={1 \over r_s} \sum_{l} \int_{{\rm BZ}}{d^2p\over (2\pi)^2} 
b^\dagger_{l}({\bf p})\left( i {\partial 
\over \partial p_y}\right) b_{l}({\bf p}),\\
&Q_J&={1 \over r_s}\sum_{l}  \int_{{\rm BZ}}{d^2p\over (2\pi)^2}
 b^\dagger_{l}({\bf p})\left[l+{1 \over 2}+{\pi}\{{{r_s}^2\left(i 
{\partial \over \partial p_x}-{p_y \over {2 \pi}}\right)^2   
+r_s^{-2}\left( i {\partial \over \partial p_y}\right)^2}\}\right]
b_{l}({\bf p}),\nonumber \\
&Q&=\sum_{l} \int_{{\rm BZ}}{d^2p\over (2\pi)^2} b^\dagger_{l}({\bf p})
b_{l}({\bf p}), \nonumber
\end{eqnarray}
where $r_s$ is the asymmetry parameter. From the above expressions it
is easy to see that magnetic translation 
operators are actually the generators of translations in the momentum space, 
and are called ${\bf K}$-transformation. $Q_X$ translates the momentum in 
$x$-direction and $Q_Y$ translates the momentum in the $y$-direction. They 
commute with Hamiltonian, 
\begin{equation}
[Q,{\cal H}]  =
[Q_{X},{\cal H}]=
[Q_{Y},{\cal H}]= 
[Q_{J},{\cal H}]= 0.
\end{equation}
So they are constants of motion of the present system. They
satisfy the algebras,
\begin{eqnarray}
&[Q_{X},Q_{Y}]& = {i \over eB}Q,\nonumber \\
&[Q_{J},Q_{X}]& = iQ_{Y}, \\
&[Q_{J},Q_{Y}]& = -iQ_{X}, \nonumber
\end{eqnarray}
and
\begin{equation}
[Q_X,Q]=
[Q_Y,Q]= 
[Q_J,Q]= 0.
\end{equation}

The first commutation relation shows that it is impossible to diagonalize 
$Q_X$ and  $Q_Y$ same time. Eigenstate of  $Q_X$ breaks  $Q_Y$-invariance. 
These conserved charges are integrals of charge densities, 
$j^{0}({\bf x})$, $j_X^{0}({\bf x})$, $j_Y^{0}({\bf x})$, $j_J^{0}
({\bf x})$, of Noether currents which are given as, 
\begin{eqnarray} 
j^{\mu}({\bf x})&=&{\rm Re}(\Psi^{\dagger} v_{\mu} \Psi),\nonumber \\
j_X^{\mu}({\bf x})&=&{\rm Re}(\Psi^{\dagger} v_{\mu}X \Psi)-
{1 \over{eB}}\delta_y^{\mu} {\cal L},\\
j_Y^{\mu}({\bf x})&=&{\rm Re}(\Psi^{\dagger} v_{\mu}Y \Psi)+{1
 \over{eB}}\delta_x^{\mu} {\cal L},\nonumber \\
j_J^{\mu}({\bf x})&=&{\rm Re}(\Psi^{\dagger} v_{\mu} J \Psi)+
\epsilon_{0{\mu}{i}} x^i {\cal L},\nonumber
\end{eqnarray}
where $v^{\mu}=(1,{\bf v})$, $J={eB \over 2}(\xi^2+\eta^2-X^2-Y^2)$, and 
${\cal L }$ is the Lagrangian density. The commutation relations between 
the above charges and the current densities read 
\begin{eqnarray}
&[Q_X,j^{_\mu}({\bf x})]&=-{i \over eB}\partial_y j^{\mu}({\bf x}),\nonumber\\
&[Q_Y,j^{_\mu}({\bf x})]&={i \over eB}\partial_x j^{\mu}({\bf x}),\nonumber\\
&[Q_X,j_X^{_\mu}({\bf x})]&=-{i \over eB}\partial_y j_X^{\mu}({\bf x}),\\
&[Q_Y,j_X^{_\mu}({\bf x})]&={i \over eB}\partial_x j_X^{\mu}({\bf x})-
{i \over eB}j^{\mu}({\bf x}),\nonumber\\
&[Q_X,j_Y^{_\mu}({\bf x})]&=-{i \over eB}\partial_y j_X^{\mu}({\bf x})+
{i \over eB}j^{\mu}({\bf x}),\nonumber\\
&[Q_Y,j_Y^{_\mu}({\bf x})]&={i \over eB}\partial_x j_Y^{\mu}({\bf x}).
\nonumber
\end{eqnarray}
Thus we see that  $Q_X$ makes a translation in $-y$-direction of 
coordinate space and $Q_Y$ makes a translation in $x$-direction of 
coordinate space. As we saw before they are the generators of translations 
in the perpendicular directions of momentum space. 
This mixing of the coordinate space and momentum space is 
caused by non-commuting coordinates, $X$ and $Y$, and is a feature of the 
quantum Hall system. 
 
As is clear from the eigenvalue, $E_l$, Landau energy gap, 
$E_{l+1}-E_l$ is proportional to the magnetic field and becomes large in 
the strong magnetic field. At low temperature, it is sufficient to study 
many-body state within one Landau levels where Fermi energy is located.  
The free Hamiltonian depends on only total electron number and is irrelevant 
if dynamics within one Landau levels are concerned. We study 
the interaction Hamiltonian within one Landau level space and obtain the 
Hall gas state. 

The interaction Hamiltonian within one Landau levels is given in the von 
Neumann lattice representation as,
\begin{equation}
{\cal H}
={1 \over 2} \int{{ {d^2 k} \over (2\pi)^2}\rho({\bf k})V({\bf k})
\rho(-{\bf k})},
\end{equation}
where the density operator is defined by the operators of the particular
Landau levels in Eq.~(37). We study the Hartree Fock approximation in the 
momentum space. It is instructive, however, to
write this Hamiltonian in the coordinate space as,
\begin{equation}
{\cal H}  ={1 \over 2} \sum_{{\bf X}_i,{\bf X}'_i} b^\dagger_{l}({\bf X}_1)
b_{l}({\bf X}_1')V({\bf X},{\bf Y},{\bf Z})b^\dagger_{l}({\bf X}_2)b_{l}
({\bf X}_2'),
\end{equation}
\begin{equation}
{\bf X}={\bf X}_1-{\bf X}_1',{\bf Y}={\bf Y}_1-{\bf Y}_1',{\bf Z}={\bf
 X}_1-{\bf X}_2'.
\end{equation}
Here the Fourier transform of the operator is defined using the
operator in the fundamental region of momentum as,
\begin{equation}
{\rm b}({\bf X})=\int_{\rm BZ} {{d^2 p} \over (2\pi)^2} {\rm b}({\bf
 p})e^{-i{\bf p}\cdot {\bf X}},
\end{equation}
where the coordinates are defined in the lattice sites. The {\bf K} 
transformation is expressed as,
\begin{eqnarray}
b_{l}({\bf p})& & \rightarrow b_{l}({\bf p+K}),\\
b_{l}({\bf X})& & \rightarrow e^{i {\bf K}{\cdot} {\bf X}}b_{l}({\bf X}),\\
\rho({\bf k})& & \rightarrow 
e^{i({a \over 2\pi} {\hat k}_x) K_y} \rho({\bf k}).
\end{eqnarray}
Delicate problems connected  with the boundary condition in momentum space 
are resolved using the gauge freedoms of operators and slight modifications are
required in the above transformations\cite{von
neumann2,stripe-hf1}. Obviously ${\cal H}$ is invariant 
under the {\bf K}-transformation. An 
ordinary free Hamiltonian has an 
momentum dependent one particle energy and is not invariant under 
{\bf K}-transformation. 
So {\bf K} symmetry is a characteristic symmetry of the 
quantum Hall system. Hall gas state has a momentum 
dependent one-particle energy and breaks {\bf K} symmetry spontaneously. 
By the Coulomb interaction, the broken state is realized as ground state. 
Hence Goldstone theorem is applied and gapless Nambu-Goldstone-mode emerges. 
The properties of Nambu-Goldstone mode and related problems will be 
studied later.
  
\subsection{Mean field solution of anisotropic Hall gas (stripes)} 

Self-consistent mean field of the form, 
\begin{equation}
U_0({\bf X - \bf X'})= \langle b_{l}^\dagger({\bf X'}) b_{l}({\bf X}) \rangle
\end{equation}
is obtained from Hartree-Fock approximation of ${\cal H}$, where 
$U_0(\bf X - \bf X')$ is unknown now and is determined later. We first 
plug the above expectation value into the Hamiltonian and obtain mean 
field Hamiltonian which is bi-linear in the field. So it is easy to 
diagonalize the mean field Hamiltonian and to find the ground state by 
filling electrons up to certain filling factor. Thus the expectation value 
of two point function is obtained and is required to agree  with the 
initial expectation value. In this way mean field solution which satisfies 
self-consistency condition is obtained. From $(\bf X -X')$ dependence of the 
expectation value, lattice translational invariance is preserved and 
one-particle energy depends upon the momentum. Momentum dependent kinetic 
energy did not exist in the free Hamiltonian Eq.~(36) but is generated 
self-consistently in the mean field Hamiltonian.  
Thus {\bf K} symmetry is broken spontaneously.

Let us obtain the Hall gas state of filling $\nu $ in the momentum space 
that breaks {\bf K} symmetry minimally i.e., $K_x$-invariant solution where 
one-particle energy depends on $p_y$ but does not depend on $p_x$ explicitly. 
Expectation value is assumed symmetric in ${p_y}$ and is given as, 
\begin{equation}
\langle b^\dagger_{l}({\bf p})b_{l}({\bf p'}) \rangle=
(2\pi)^2 {\delta({\bf p}-{\bf p'})}{\theta(\pi \nu' -|p_y|)},
\end{equation}
where $\nu'=\nu-l$. 
This state preserves $K_x$-symmetry and breaks $K_y$-symmetry 
spontaneously. The mean field Hamiltonian is expressed as,
\begin{equation}
{\cal H}_m=\int_{{\rm BZ}}{d^2 p\over (2\pi)^2}{\epsilon}_l({\bf p})
 b^\dagger_{l}({\bf p})b_{l}({\bf p})+E_l(0),
\end{equation}
where $\epsilon({\bf p})$ is one-particle energy and $E_l(0)$ is zero-point 
energy. Let us assume that $ \mu_0({\nu})$ to be the chemical potential. 
The self-consistency condition is given by, 
\begin{eqnarray}
\epsilon({\bf p})
&=&
\int_{{\rm BZ}}{d^2 p'\over (2\pi)^2}v_{\rm HF}({\bf p'-p})
\theta(\mu_0(\nu)-{\epsilon}({\bf p'})),\\
\nu'
&=&
\int_{\rm BZ} {d^2 p'\over (2\pi)^2}
\theta[\mu_0(\nu) - {\epsilon}( {\bf p'} )].
\end{eqnarray} 
In the above equations $v_{\rm HF}({\bf p'-p})$ is the Hartree-Fock
potential energy that combine the Coulomb potential with mean field as, 
\begin{equation}
v_{\rm HF}({\bf p})=\sum_{\bf N}\{v_l(2\pi{\tilde{\bf N}})e^{i{\bf p}
\times{\bf N}}-v_l(2\pi{\tilde{\bf N}}-{\tilde{\bf p}})\},
\end{equation}
where $v_l(k)=e^{-k^2/4\pi}[L_l(k^2/4\pi)]^2 V(k)$. 
By solving self-consistency conditions, 
$\epsilon({\bf p})$ and  $E_l(0)$ are obtained. $p_y$ dependent 
one-particle energy thus emerges. One-particle energy at half filling is 
given in Fig.~14. As is seen in Fig.~15, Fermi surface is parallel to 
$p_x$ axis and $K_x$ symmetry is preserved. The asymmetry parameter
$r_s$ is determined by minimizing the total energy.
%
%
\begin{figure}[th]
\centerline{\includegraphics[width=9cm]{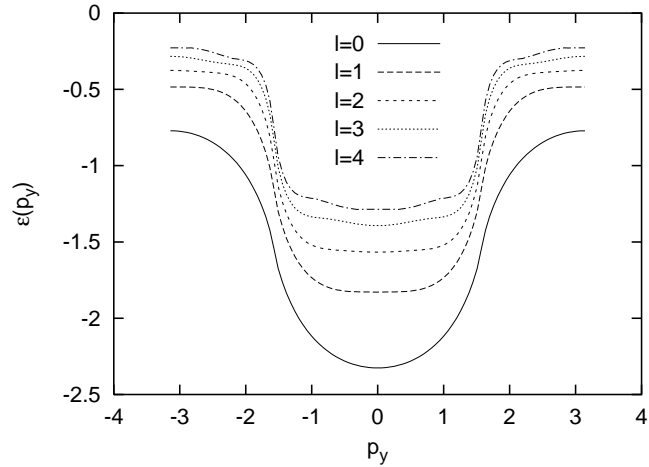}}
\vspace*{8pt}
\caption{One-particle energy spectrum of anisotropic Hall gas at the 
filling factor ${\nu}=l+{1 \over 2 }$. The unit of energy is 
${q^2 \over a}$ 
}
\end{figure}
%
%
\begin{figure}[th]
\centerline{\includegraphics[width=7cm]{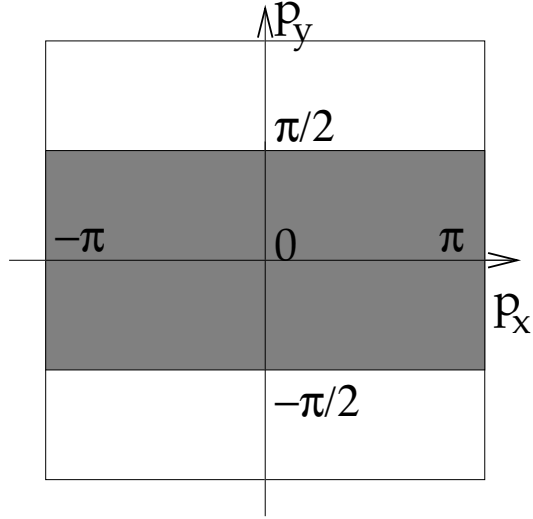}}
\vspace*{8pt}
\caption{Fermi surface of anisotropic Hall gas is parallel to $p_x$ axis.}
\end{figure}

Because the Coulomb potential decreases slowly at infinity, slope 
of one-particle energy at the Fermi energy, Fermi velocity, diverges. 
If the Coulomb potential is screened and screened Coulomb potential is used, 
the Fermi velocity becomes finite. 

From the total energy we compute physical quantities that 
show thermodynamic properties of this many-body state. Since the kinetic energy
is produced by interaction only, it is expected that the present Hall gas 
reveals peculiar properties that are different from those of ordinary electron 
gas. Pressure and compressibility of Hall gas with neutralizing uniform 
background charge are computed and are given in 
Fig.~16\cite{thermodynamical}. As is seen in Fig.~16, pressure becomes 
negative. Negative pressure means that the gas has a tendency to shrink by 
itself. Due to the neutralizing background charge, the magnitude of size of 
shrink is normally very small. But it has important effects for the IQHE 
in realistic systems. Implications to metrology will be discussed in 
Section 6. 
%
%
\begin{figure}[th]
\centerline{\includegraphics[width=9cm]{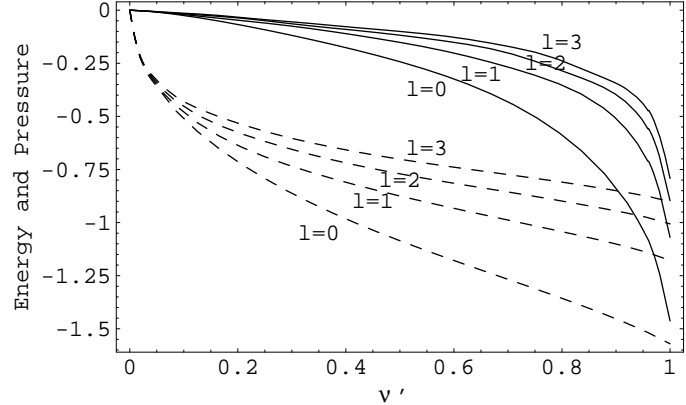}}
\vspace*{8pt}
\caption{Energy per particle and pressure of anisotropic Hall gas as a 
function of the effective filling factor ${\nu'}$ in intra Landau 
levels for the filling factor ${\nu}=l+{\nu'}$.  The solid lines show the 
pressure in the unit of ${q^2 \over a^3}$ and the dashed lines show the 
energy in the unit of ${q^2 \over a}$.
}
\end{figure}
%
%
\begin{figure}[th]
\centerline{\includegraphics[width=9cm]{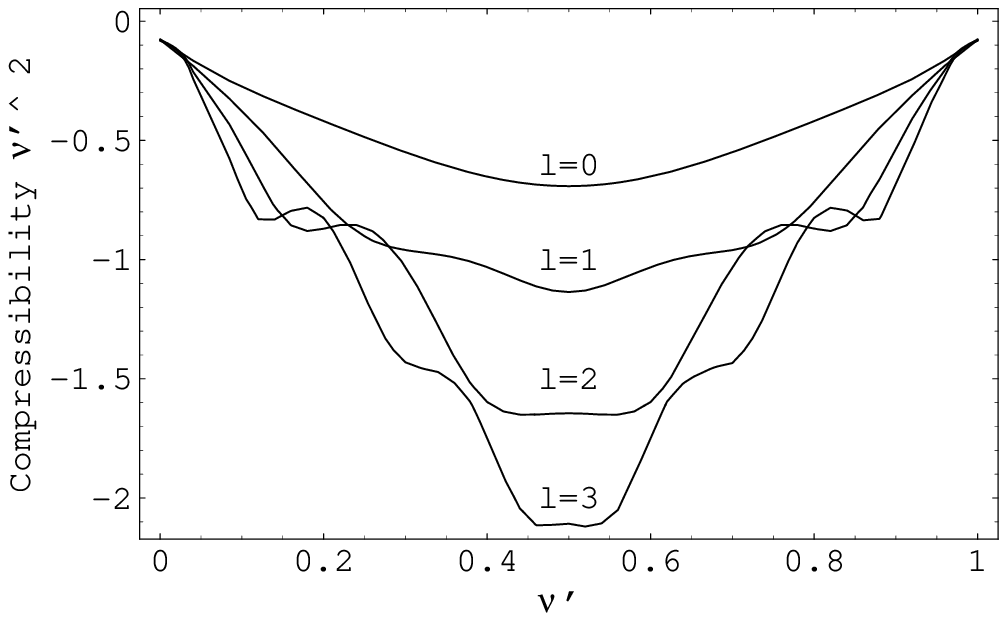}}
\centerline{\includegraphics[width=9cm]{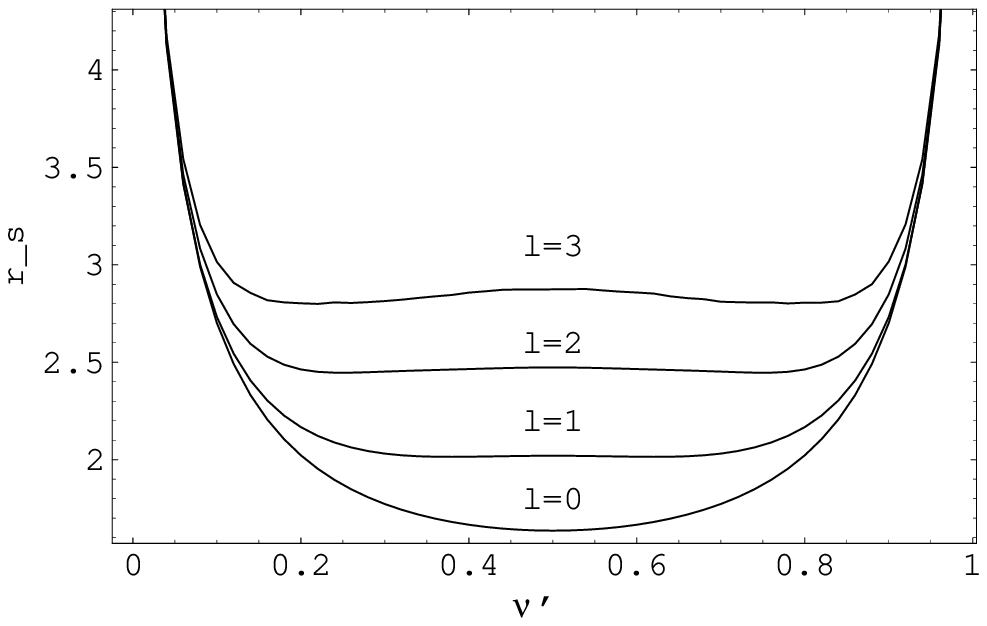}}
\vspace*{8pt}
\caption{ Compressibility times ${\nu'}^2$ of anisotropic Hall gas in 
the unit of ${a^3 \over q^2}$ and the value of asymmetry parameter $r_s$ as a 
function of the effective filling factor ${\nu'}$ in intra Landau levels 
for the filling factor ${\nu}=l+{\nu'}$.   
}
\end{figure}

Compressibility becomes also negative and is given in Fig.~17. 
In ordinary electron gas, compressibility becomes negative only in low 
density where interaction effects surpass free kinetic energy. In the Hall 
gas, ordinary kinetic energy is frozen by the magnetic field so interaction 
effect is enhanced and this phenomenon occurs in arbitrary density. The
asymmetry parameter $r_s$ changes slightly with the filling factor as
is shown in Fig.~17.

\subsection{Charge density and current density profile} 

Asymmetric Fermi surface in the momentum space means that the 
orientational symmetry is broken. To see a direct signal of 
orientational symmetry breaking, we study the density in coordinate
space. Number density and current density in real 
space become  also asymmetric and are given by, 
\begin{eqnarray}
\langle \rho ({\bf x}) \rangle 
&=& 
\int{d^2k \over (2\pi)^2} e^{i\bf k \cdot \bf x }
\int_{\rm BZ} {d^2p\over (2\pi)^2}
\langle b^\dagger_{l}({\bf p})b_{l}({\bf p}+a{\hat {\bf k}}) \rangle
\langle f_l|e^{-i(k_x \xi+k_y \eta)}|f_l \rangle \nonumber 
\\ 
& &\times e^{ i{a \over 4\pi} {\hat k_x}(2p_y+a{\hat k}_y)},\\
\langle {\bf j} ({\bf x}) \rangle 
&=& 
\int{d^2k\over (2\pi)^2} 
e^{i\bf k \cdot \bf x }
\int_{{\rm BZ}}{d^2p\over (2\pi)^2}
\langle b^\dagger_{l}({\bf p})b_{l}({\bf p}+a{\hat {\bf k}}) \rangle
\langle f_l|{1 \over 2}\{{\bf v},e^{-i(k_x \xi+k_y \eta)}\}|f_l \rangle 
\nonumber \\  
& &
\times e^{ i{a \over 4\pi }  {\hat k_x} (2p_y+a{\hat k}_y)}.
\end{eqnarray}
The {\bf phase factor} in the above equation is due to commutation relation 
between the guiding center coordinates, $X$ and $Y$, and causes particle's 
motion toward  perpendicular to electric field. In the present equation, 
it leads the density profile in coordinate space to be uniform in 
$y$-direction and is periodic in $x$-direction with the 
period $r_s a$, Fig.~18. Thus the density 
in the real space is uniform in the direction orthogonal to Fermi surface 
in the momentum space. So the present anisotropic Hall gas is equivalent to 
unidirectional charge density wave of Koulakov, Fogler, and 
Shklovskii\cite{stripe1} and of Moessner and Chalker\cite{stripe2}.

%
%
\begin{figure}[th]
\centerline{\includegraphics[width=10cm]{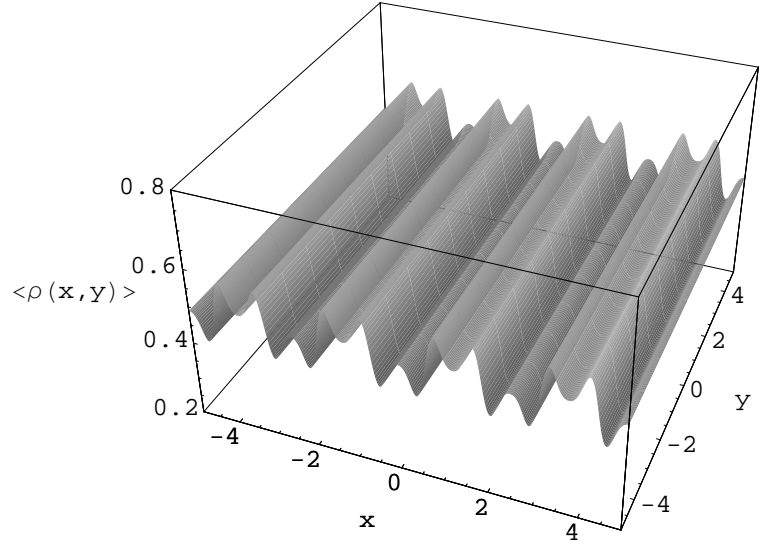}}
\vspace*{8pt}
\caption{Density profile in the coordinate space of anisotropic Hall gas at 
$\nu = 2+1/2$. The unit of density is ${1 \over a^2 }$ and the unit of
 coordinates is $a$.}
\end{figure}
 
Number density and current density of the present mean field are 
given in Fig.~19. The current density in the real space is perpendicular to 
the gradient of density. Current flows locally in the direction of the 
uniform density, but the total current vanishes. 

From the shape of Fermi surface, all the one-particle states are filled in 
the $p_x$ direction. Unfilled one-particle state of the lowest energy in 
the $p_x$ direction is in the next Landau levels, which has a large energy 
gap in the system of the strong magnetic field. Hence the state behaves 
like the integer quantum Hall state against the perturbation in this 
direction. Resistance in this direction vanishes. 
 
%
%
\begin{figure}[th]
\centerline{\includegraphics[width=8cm]{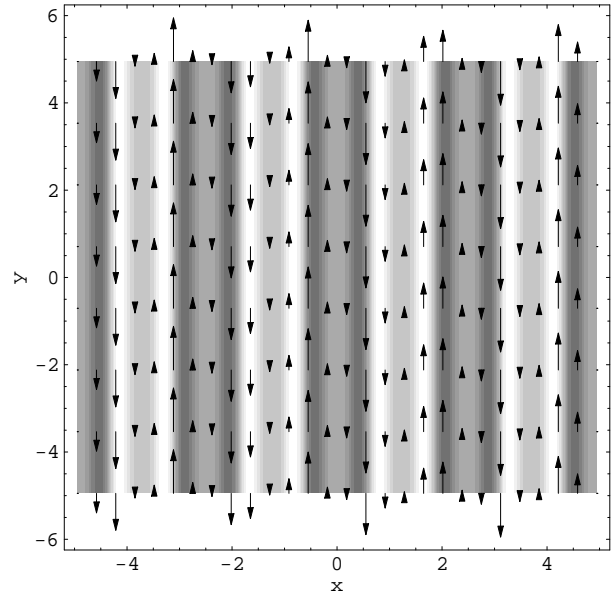}}
\vspace*{8pt}
\caption{Current profile in the coordinate space of anisotropic Hall gas at 
$\nu = 2+1/2$. Arrows show current vector and shadows show low  density
 area. The unit of coordinates is $a$.
}
\end{figure}

\subsection{
\bf Symmetry breaking and Nambu-Goldstone zero-mode}     

Translational and orientational symmetry are broken in the anisotropic Hall
gas obtained in the previous section. So {\bf Nambu-Goldstone zero mode} 
appears. Center coordinates $(X,Y)$ are non-commuting and magnetic 
translations are non-commutative. Consequently  Goldstone theorem is 
nontrivial and spectrum of Nambu-Goldstone zero mode would reflect these 
non-commutative natures\cite{neutral mode}. 
From both of the Fermi surface, Fig.~15, and the charge density, Fig.~18, 
we see that  $Q_Y$ and $ Q_J$ are broken spontaneously.

Since the conserved charges $Q_Y$ and $ Q_J$ are broken spontaneously, 
Nambu-Goldstone mode appears. Proof for the theorem is given in the 
following using the above commutation relations. Let the broken state be 
the eigenstate of the Hamiltonian, a unitary operator that corresponds 
to $Q_X$, the unitary operator that translates the system by $r_s a$ in 
$x$-direction, and the total charge simultaneously, 
\begin{eqnarray}
{\cal H}|n \rangle &=&E_n|n \rangle,\\
e^{i2 \pi Q_X/({r_s L_x})}|n \rangle &=& e^{i2 \pi Q_X^{(n)}/({r_sL_x})}
|n \rangle, \\
e^{i 2\pi r_s Q_Y/a}|n \rangle &=& e^{i2 \pi r_s Q_Y^{(n)}/a} |n \rangle, \\
Q |n \rangle &=& N_e |n \rangle,
\end{eqnarray}
where  the states including vacuum, $|0 \rangle$, is periodic in $x$-direction 
with the periodicity, $r_s a$, and $L_x$ is the length of the system in 
$x$-direction. 

By taking expectation value of the algebra, a state of a non-uniform charge 
density or equivalently a state of partially filled state with a definite 
Fermi surface  leads that the symmetry is broken spontaneously. 
Let us use the algebra, Eq.~(75), in the coordinate space and its expectation 
value, 
\begin{equation}
{i \over eB } \partial_x \langle j^0({\bf r},t) \rangle
=\langle 0|[Q_Y,j^0({\bf r},t)]|0\rangle
=\int{d{\bf r'}\langle 0|[j_Y^0({\bf r'},t'),j^0({ \bf r},t)]}|0\rangle
\label{order}
\end{equation}
To rewrite the right-hand side, we use 
\begin{equation}
e^{i eB \Delta y Q_X}j_Y^0({\bf r})e^{-i eB \Delta y Q_X}
=j_Y^0[{ \bf r}+(0,\Delta y)]-\Delta y j^0[{\bf r}+(0,\Delta y)],
\end{equation}
and decompose the coordinate $x$ into $x=N_x r_s a+{\bar x }$, where $N_x$
is an integer and $0 \leq{{\bar x}} \leq r_s a$. Using the commutation 
relation, the definition of the broken state, and the above coordinate 
expression, the right-hand side of Eq.~(\ref{order}) is written as,
\begin{eqnarray}
& &
a^3\sum_{n\ge 0} \int_0^{r_s a} {d {\bar x'} \over r_s}
\left[ \left( \langle 0|j_Y^0[({\bar x'},0),0]
|n \rangle \langle n|j^0[({\bar x},0),0]|0 \rangle e^{i(t'-t)(E_n-E_0)}-
H.C.\right)\right.\nonumber \\
& &
\times\delta({Q_X}^{(0)}-{Q_X}^{(n)}) 
\delta_{1/{r_s}}({Q_Y}^{(0)}-{Q_Y}^{(n)})+
{1 \over eB i} \left( \langle 0|j^0[({\bar x'},0),0]|n \rangle 
\langle n|j^0[({\bar x},0),0] |0 \rangle\right.\nonumber \\ 
& &
\left.\left.\times e^{i(t'-t)(E_n-E_0)}+H.C. \right)
\delta'({Q_X}^{(0)}-{Q_X}^{(n)})
\delta_{a/{r_s}}({Q_Y}^{(0)}-{Q_Y}^{(n)}) 
\right],
\end{eqnarray}
where $\delta_{a/{r_s}}(Q_Y)$ stands for the delta function with the
period
${a\over r_s}$. Because the left-hand side does not depend on $t$, the 
energy gap, ${\Delta} E_{NG}({\bf q})=E_n-E_0$, where $q$ is the momentum of 
$|n\rangle$, vanishes in the small $q$ limit. 
Thus the existence of gapless excitation mode is proved. 

Properties of Nambu-Goldstone mode is complicated because the center 
coordinates are non-commuting and magnetic translations are also 
non-commutative. To calculate its spectrum, we apply single mode 
approximations  which were successfully used to study the properties of 
liquid-helium\cite{sma-feynman} and fractional quantum Hall 
states\cite{sma-fqh}. 

Let us map the density operator and Hamiltonian into $l$-th Landau level 
space as, 
\begin{eqnarray} 
\rho_*({\bf k})&=&P_l \rho ({\bf k}) P_l \nonumber \\
&=&\int_{{\rm BZ}}{d^2p\over (2\pi)^2}
b^\dagger_{l}({\bf p})b_{l}({\bf p}+a{\hat{\bf k}}) 
e^{ i{a \over 4\pi} {\hat k}_x(2p_y+a{\hat k}_y)}, \\
H^{(l)}&=&
{1 \over 2} \int {d^2k\over (2\pi)^2}
\rho_*({\bf k}) v_l(k)\rho_*({-\bf k}),
\end{eqnarray}  
where $v_l(k)=e^{-k^2/{4 \pi}}[L_l(k^2/{4 \pi})]^2 2 \pi q^2/k$. 
This density operator satisfies the commutation relations,
\begin{equation} 
[\rho_*({\bf k}),\rho_*({\bf k'})]
=-2i{\sin}\left({\bf k} \times {\bf k'}\over 4\pi\right)
\rho_*({\bf k+\bf k'}),
\end{equation} 
\begin{equation} 
[Q_X,\rho_*({\bf k})]={k_y \over 2 \pi}\rho_*({\bf k}),
\end{equation}
\begin{equation} 
[Q_Y,\rho_*({\bf k})]=-{k_x \over 2 \pi}\rho_*({\bf k}).
\end{equation}
The state defined by 
\begin{equation}
|{\bf k} {\rangle}= \rho_*({\bf k})|0 {\rangle}
\end{equation}
has a quantum number of Nambu-Goldstone mode and couples with the 
corresponding current. So we assume that this state is one particle
state of Nambu-Goldstone mode and computes 
its energy from the expectation value,
\begin{equation}   
\Delta({\bf k})={\langle{\bf k} |(H^{(l)}-E_0)|{\bf k} \rangle \over \langle
{\bf k}|{\bf k}\rangle}.
\end{equation}

To compute the numerator we use,
\begin{eqnarray} 
{\langle{\bf k} |(H^{(l)}-E_0)|{\bf k} \rangle}
&=&\langle 0 |[\rho_*(-{\bf k}),H^{(l)}] \rho_*({\bf k}) | 0 \rangle
\nonumber
\\
&=&-\langle 0 |[H^{(l)},\rho_*({\bf k})] \rho_*(-{\bf k}) | 0 \rangle,
\end{eqnarray}
which are obtained from reflection symmetry,
\begin{eqnarray} 
\langle 0|\rho_*({\bf k}) \rho_*(-{\bf k})|0 \rangle
&=& \langle 0|\rho_*(-{\bf k}) \rho_*({\bf k})|0 \rangle,\\
\langle 0|\rho_*({\bf k}) H^{(l)} \rho_*(-{\bf k})| 0 \rangle
&=& \langle 0|\rho_*(-{\bf k}) H^{(l)} \rho_*({\bf k})|0 \rangle,
\end{eqnarray}
and $ H^{(l)} | 0 \rangle = E_0 |0 \rangle$.

Using commutation relations and the Hamiltonian, we find the spectrum from 
single mode approximation as, 
\begin{equation}
\Delta({\bf k})=|k_y|\{Ak_x^2+Bk_y^4+O(k_x^2 k_y^2,k_x^4)
{\rm ln}\vert k_y\vert\},
\end{equation}
where A and B are constants.
The energy spectrum is anisotropic and vanishes with higher powers in the 
small momentum region. At $k_x=0$ the energy depends on fifth power of $k_y$ 
and at a finite $k_x$ the energy depends linearly on the magnitude of
$k_y$. At  $k_y=0$, not only the energy vanishes exactly, but also the state 
$|{\bf k} {\rangle}$ vanishes. This is trivial from the shape of the 
Fermi surface, Fig.~15. Spectrum in wide range of ${\bf k}$ is given in 
Fig.~20.

%
%
\begin{figure}[th]
\centerline{\includegraphics[width=7cm]{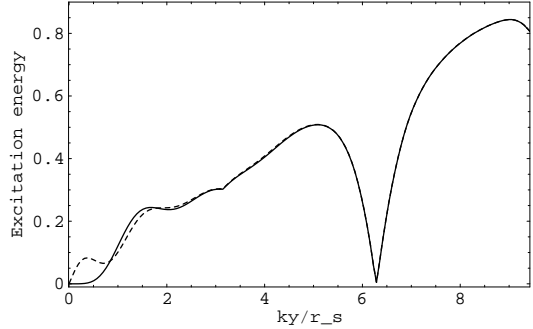}}
\vspace*{8pt}
\caption{Energy spectrum of excited states of anisotropic Hall gas at 
${\nu=2+{1 \over 2}}$ as a function of momentum $k_y$. 
The unit of ${\bf k}$ is $a^{-1}$ and the unit of the energy is 
${q^2 \over  a}$. The solid line is for  $k_x=0$ and the dashed line is 
for $k_x=1$
}
\end{figure}

\subsection{
\bf Preferred orientation under external density modulation}

Because orientational symmetry is spontaneously broken in the 
present mean field, by an infinitesimal perturbative term which breaks 
orientational symmetry, the system is forced to choose one  particular 
direction. This direction is kept even in the limit of vanishing perturbative 
term. We study which direction is chosen in the presence of small 
external density modulation in this section\cite{optimal1,optimal2}. 
Our finding for the energy of the system with external density modulation 
is counterintuitive. Although it is expected that the energy of 
stripe becomes minimum when the stripe is aligned parallel to the external 
density modulation, we found that this depends on the magnitude and 
the wave length of the density modulation. In the case of  weak and long 
wave length modulation, the energy of stripe becomes minimum when the 
stripe is perpendicular  to the external density modulation. 

We add small density modulation term of the coupling strength, $g$, and 
wave vector, ${\bf K}_{\rm ext}$, into the Hamiltonian, 
${\cal H}_{\rm total}={\cal H}+{\cal H}_{\rm ext}$, 
where ${\cal H}_{\rm ext}$ is given by, 
\begin{equation}
{\cal H}_{\rm ext}=g\int{d^2 r} \rho({\bf r})\cos({\bf K}_{\rm ext}\cdot
{ \bf r}).
\end{equation}
We study the stripe state from the total Hamiltonian. For small $g$ case, 
perturbative expansion with respect to $g$ may be a good approximation. 
So first we compute the energy  correction of ground state up to the square 
of $g$ based on perturbative expansion. This energy becomes a function of 
the angle between the stripe and external modulation, ${\theta}$, and of 
the magnitude of ${\bf K}_{\rm ext}$ and is given in Fig.~21. 
%
%
\begin{figure}[th]
\centerline{\includegraphics[width=10cm]{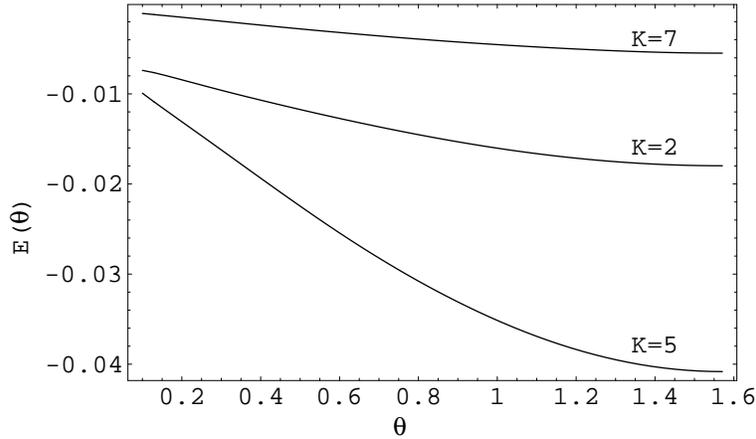}}
\vspace*{8pt}
\caption{Energy gain of the  anisotropic Hall gas as a function of the angle 
between the external periodic density modulation and intrinsic periodic 
density. The unit of the energy is ${g^2 a \over q^2}$. 
}
\end{figure}

The energy becomes minimum when the angle is ${\pi \over 2}$. 
Thus the stripe is aligned perpendicular to the external modulation. 
This result is understandable from the shape of Fermi surface. 
All the one-particle states are filled in the $p_x$ direction and unfilled 
one-particle state of the lowest energy in the $p_x$ direction is in the 
next Landau levels, which has a large energy gap in the system of the strong 
magnetic field. Hence the many-body state is hard and does not get any 
perturbative energy if the external perturbation is in the $x$-direction. 
This state behaves like the integer quantum Hall state against the 
perturbation in this direction.  Contrary to the $x$-direction, 
there are states around the Fermi surface without energy gap and the 
many-body state is soft and gets a perturbative energy if the external 
perturbation is in the $y$-direction. Thus the orthogonality between the 
external density modulation and the intrinsic density profile is caused 
by the shape of Fermi surface, Fig.~15, and is a characteristic feature of 
the quantum Hall system. 
 
If ${\bf K}_{\rm ext}$ is equal to the intrinsic wave 
vector of the stripe, a HF solution of ${\cal H}$ becomes a HF solution of 
total Hamiltonian. Both directions are the same in this case. So the state 
realized in the present case is different from the state in the previous 
case. From comparison of the energies of the both states  we obtain phase 
diagram as a function of several parameters, Fig.~22.
%
%
\begin{figure}[th]
\centerline{\includegraphics[width=13cm]{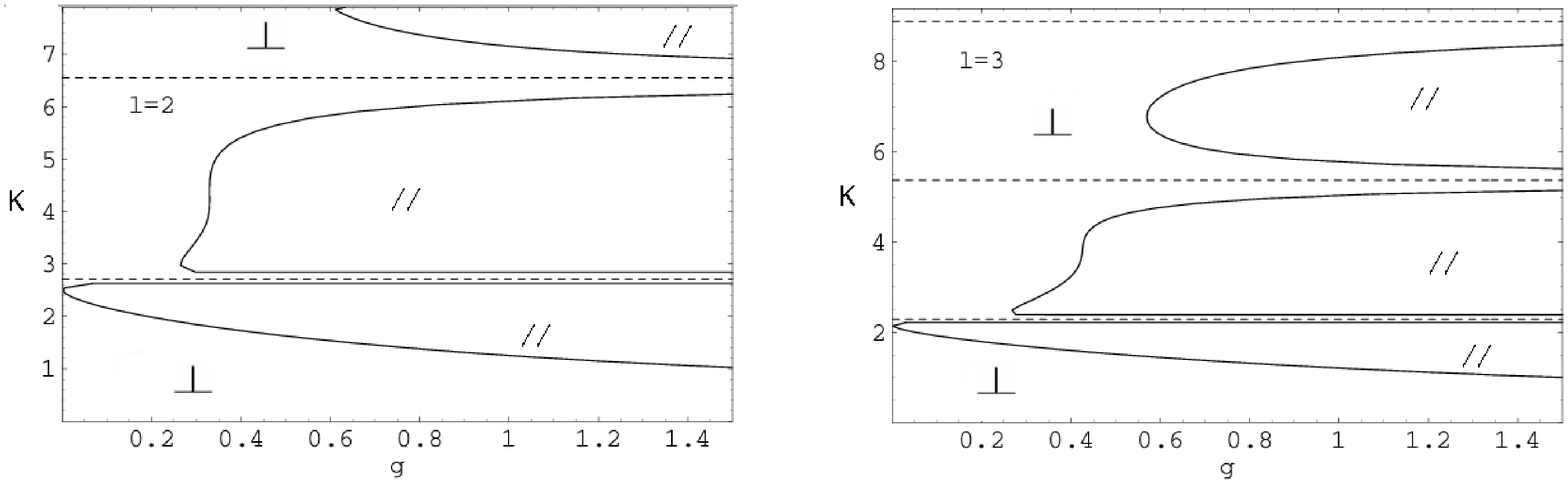}}
\vspace*{8pt}
\caption{Phase diagram of parallel phase and orthogonal phase for 
$l$=2 and 3.}
\end{figure}

In small $g$, {\bf orthogonal phase} in which both direction are orthogonal 
is realized and in large $g$, {\bf parallel phase} in which both directions 
are parallel is realized\cite{moduration1,moduration2}. In the long wave 
length limit, the {\bf orthogonal phase} is always realized. Systems with
in-plane magnetic field in $x$-direction is expressed with the vector 
potential, 
${\bf A}+{\bf A}_{\rm in}$, ${\bf A}={B \over 2}(y,-x,0)$, 
${\bf A}_{\rm in}=(0,0,B_{\rm in}({\sin y K \over K}))$. 
In the limit $K \rightarrow 0$,
in-plane magnetic field becomes uniform, and the
orthogonal phase, in which the Fermi surface is parallel to $p_x$
direction is realized. Thus the effect of the in-plane
constant magnetic field is almost the same as the effect of 
external modulation studied in this section. The orthogonal phase is
realized. This is consistent with experiments\cite{in-plane}.

\subsection{Electronic transport  }

{\bf Hall conductance}: 
We study the electric transport of the system in  which the 
anisotropic Hall gas fills whole spatial regions in this section. 
Using the energy eigenvalue of Section 3.2, $\varepsilon_l({\bf p})$,  
the propagator for ${\tilde b}_l({\bf p})$ is expressed as, 
\begin{eqnarray}
{\tilde S}^{(c)}_{l_1l_2}(p)&=&U_{l_1l_1'}({\bf p}) S^{(c)}_{l_1'l_2'}(p) 
U^\dagger_{l_2'l_2}({\bf p})\nonumber \\
 S^{(c)}_{l_1l_2}(p)&=& {\delta_{l_1l_2}\over p_0-(E_{l_1}+
\varepsilon_{l_1}({\bf p}))}
\label{prop}
\end{eqnarray}
We substitute  Eq.~(\ref{prop}) to Eq.~(\ref{topological}), and we have 
\begin{equation}
\sigma_{xy}={e^2\over h}(l+\nu '),
\end{equation}
at $\nu=l+\nu'$. 
Thus the Hall conductance is unquantized and is proportional to the filling 
factor in the Hall gas regime, where the many-body state of this Fermi 
energy is anisotropic quantum Hall gas. 

{\bf Longitudinal resistances}: 
The $K_x$-invariant anisotropic Hall gas states at $\nu=l+1/2$ have the Fermi 
surface parallel to $p_x$ axis. 
In $p_x$ direction, there is no empty state within the same Landau level. 
The lowest energy unoccupied state is in the next Landau level and has 
the cyclotron energy gap. This direction is like the integer quantum 
Hall state. Thus, longitudinal conductance in this direction vanishes, 
\begin{equation}
\sigma_{xx}=0.
\end{equation}

In the $p_y$-direction, there are states in the same Landau levels which 
have no energy gap. The system behaves like a bunch of parallel 
one-dimensional systems in coordinate space. Each part is like a narrow 
one-dimensional strip of high density and of finite current density. 
If the small voltage contact is attached to one of one-dimensional systems 
\cite{one-dimensional} as in Fig.~23, the conductance in this situation
is computed from Landauer 
formula. The current in this region is carried by one-dimensional electrons 
which have left-moving modes and right-moving modes in narrow spatial region. 
So the situation is different from the edge states of the bulk quantum Hall 
systems where only one chiral mode exists in one edge.  
%
%
\begin{figure}[th]
\centerline{\includegraphics[width=5cm]{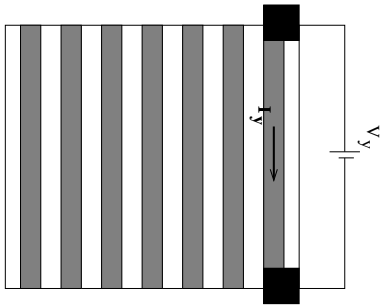}}
\vspace*{8pt}
\caption{A schematic illustration of a measurement of the longitudinal 
conductance.}
\end{figure}

In Landauer formula\cite{landauer} to this one-dimensional system, small 
difference of the chemical potential, the level density, and the velocity 
are used. The velocity is given as
\begin{equation}
v_y={\partial \varepsilon(p_y)\over\partial p_y},
\label{v}
\end{equation}
and number of states of the extended states in the chemical potential 
difference $\Delta\epsilon$ is,
\begin{equation}
\Delta n={1\over2\pi}{\partial p_y\over\partial\varepsilon}
\Delta\varepsilon (1+C),
\label{C}
\end{equation}
where $C=(\partial\delta_p/\partial p)_{p_F}/L$ is the correction term 
which comes from the phase shift $\delta_p$ due to impurity scattering, 
and $L$ is the length of the one-dimensional system. 
Combining Eqs.~(\ref{v}) and (\ref{C}), we have total current 
\begin{eqnarray}
I_y&=&ev_y\Delta n\nonumber\\
&=&{e^2\over2\pi}(1+C)V_y,
\end{eqnarray}
where $V_y$ is voltage in $y$-direction and  the chemical potential difference
$\Delta\epsilon$ is replaced with $eV_y$. 
The precise value of the correction term $C$ is not known. 
It is, however, expected that $C\rightarrow-1$ when the Landau level 
index goes to infinity from the scaling theory of Anderson localization 
in the system of zero magnetic field, and $C$ vanishes in the 
ideal system with no impurities. 
In ideal systems without impurities, the ideal conductance
\begin{equation}
\sigma_{yy}^0={e^2\over h},
\end{equation}
is obtained. In general cases, $\sigma_{yy}$ depends on the
disorders and may be different from $\sigma_{yy}^0$. 

Because the resistance tensor is the inverse of the conductance 
tensor,the resistance in $y$ direction vanishes and the resistance in $x$ 
direction becomes finite value. They are given as, 
\begin{equation}
R_{yy}=0,\quad R_{xx}={-\sigma_{yy} \over \sigma_{xy}\sigma_{yx}}.
\end{equation}
Thus in the measurement of resistance of the present anisotropic Hall
gas, the easy direction is the $y$-direction. The density is uniform and 
the Fermi surface is orthogonal to this direction. In the system of 
 $y$ dependent density modulation of the long wave length  and in the
 system of the in-plane magnetic field of $x$-direction, the easy
 direction becomes $y$-direction. This agrees with experiments. 

\section{
\bf Implications of gas of negative pressure }

In the previous section we have found that anisotropic quantum Hall gas 
has unusual thermodynamic properties such as negative pressure and negative 
compressibility. Due to these properties, the quantum Hall gas shrinks. 
Electron density becomes nonuniform and charge neutrality is violated 
locally. States of high density part becomes strip-like shape in a simple 
case. Shape may become more complicated in ordinary cases but in 
fact the shape is irrelevant to physical effects. So we call this non-uniform 
state strip and study its implications. In systems of disorders non-uniform 
charge state is also created. Physical effects of this non-uniform state is 
equivalent to the strip due to interactions. So there are two mechanisms of 
generating strip. One is by interaction and the other is by disorders. We 
analyze both mechanisms. 

\subsection{
Shape of the anisotropic quantum Hall gas with a negative pressure}

We show first that gas of negative pressure shrinks and is stabilized 
at a certain smaller size than the original size of the electron system. 
Compressible many-electron states of negative pressure is deformed due to 
negative pressure effects. We study uniform Hall gas composed of electrons 
and of uniform neutralizing  background charges $\rho_0$, which are composed 
of ions, first. The electrons can move but the ions can not. So when 
electrons shrink, charge neutrality is violated partly and the energy of 
the system is increased by a static Coulomb energy. The Hamiltonian of this 
system then is given as,
\begin{eqnarray}
{\cal H}&=&{1 \over 2}\int{d^2 x}{d^2 y}\;(\rho({\bf x})-\rho_0)\;v({\bf
 x-y})\;(\rho({\bf y})-\rho_0) \nonumber \\
&=&{1 \over 2}\int{d^2 x}{d^2 y} \left(\langle\rho({\bf x})\rangle-
\rho_0+:\rho({\bf x}):\right)v({\bf x-y})\left(\langle\rho({\bf y})
\rangle-\rho_0+:\rho({\bf y}):\right). 
\end{eqnarray}
The right hand side of the above equation is expanded as,
\begin{eqnarray}
{\cal H}
&=&
{1 \over 2}\int{d^2 x}{d^2 y} \left(\langle 
\rho({\bf x})\rangle-\rho_0\right)
v({\bf x-y})\left(\langle \rho({\bf y})\rangle-\rho_0\right) \nonumber \\
&+&
{1 \over 2} \int{d^2 x}{d^2 y} \left(\langle \rho({\bf x})\rangle -
\rho_0\right)v({\bf x-y}):\rho({\bf y}):   \\
&+&
{1 \over 2} \int{d^2 x}{d^2 y} 
:\rho({\bf x}):v({\bf x-y})\left(\langle\rho({\bf
                            y})\rangle-\rho_0\right) \nonumber \\ 
&+&
{1 \over 2} \int{d^2 x}{d^2 y} :\rho({\bf x}):v({\bf x-y}):\rho({\bf y}):.
\nonumber
\end{eqnarray}
The first term in the right hand side is the static Coulomb energy and 
vanish when the charge neutrality is preserved. We study the system in 
which the charge neutrality is violated in a macroscopic scale. The second 
and third terms are proportional to the normal ordered products of the 
density. Hence the expectation values of the second and third term vanish 
when we study the system within Hartree-Fock approximation. From the 
assumption that the charge distribution becomes non-uniform in macroscopic 
size, the HF approximation of the previous section is applied to the present 
system. Hence we  ignore the second and third terms and we apply the result 
of the last section to the last term  in the above Hamiltonian. 

We study first the static Coulomb energy in the right hand side. 
We compute the Coulomb energy per particle of the electron system with the 
initial value of width in the $x$-direction, $L_w$, and infinite length in the 
$y$-direction, and of the filling factor $l+\nu'$, when the gas shrinks 
uniformly by $\delta_x$ in the width as is in Fig.~24. The static Coulomb 
energy per particle of this situation is found to have the form:
\begin{equation}
E_{\rm Coulomb}=\nu' {L_w\over a} g\left({\delta_x\over L_w}
\right){q^2\over a},
\end{equation}
where $g(x)$ is a monotonically increasing function which 
behaves as $-{x^2\over2}\log x$ in $x\rightarrow 0$ and diverges at $x=1$ 
logarithmically. The HF energy is a decreasing function of $x$ due to 
negative pressure effect. The total energy is expressed as a function of 
the $\delta_x$ in Fig.~25. 
The total energy becomes minimum at one value of the $\delta_x$. 
At this value the Hall gas is stabilized. The pressure vanishes 
and the compressibility becomes positive. This  value depends on the filling 
factor. $\delta_x$ nearly vanishes when the filling of the Hall gas is 
large, and $\delta_x$ becomes finite if the filling is small. 
Hence the Hall gas shrinks substantially and is deformed to a smaller
size if the filling is slightly away from integer. 
%
%
\begin{figure}[th]
\centerline{\includegraphics[width=8cm]{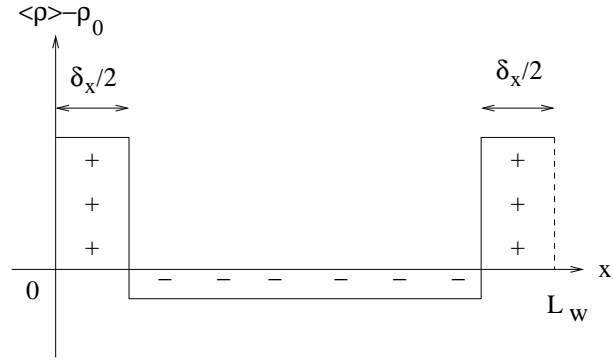}}
\vspace*{8pt}
\caption{Electron density of anisotropic Hall gas when the charge 
neutrality is violated slightly.
}
\end{figure}

%
%
\begin{figure}[th]
\centerline{\includegraphics[width=7cm]{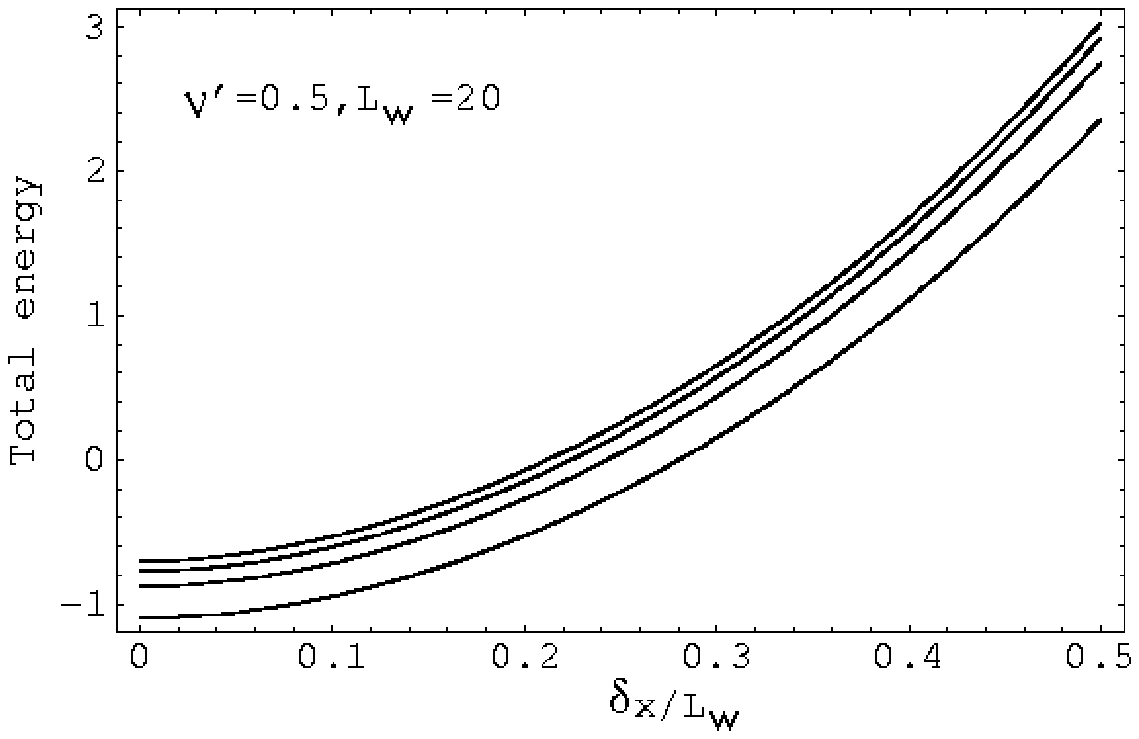}}
\centerline{\includegraphics[width=7cm]{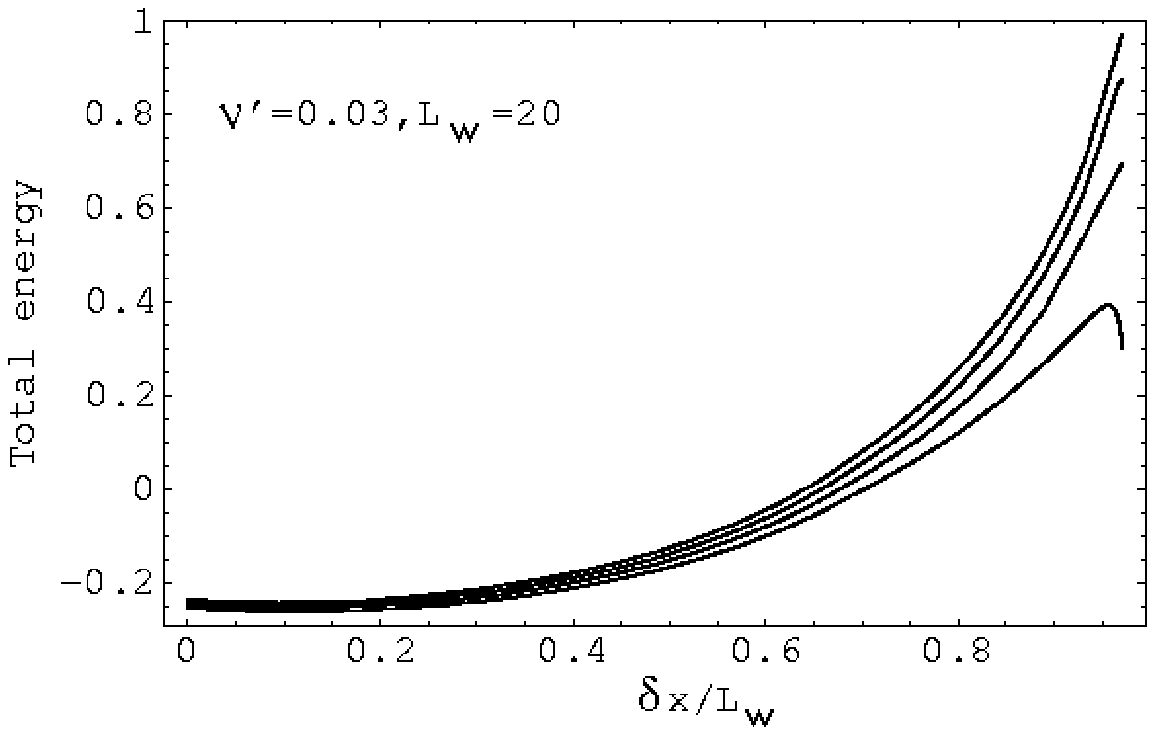}}
\vspace*{8pt}
\caption{The total energy of anisotropic Hall gas at $l=0,1,2$, and 3 
in the order from below when the charge neutrality is violated slightly 
as a function of ${\delta_x \over L_w}$. The width of strip is 
$L_w=20a$ and ${\nu'=0.5}$ and  ${\nu'=0.03}$. The unit of energy is 
${q^2 \over a}$  
}
\end{figure}

\subsection{Interplay between localization length and finite size}

In real systems of GaAs, there are long range disorder potentials. 
Localized states due to disorders have discrete energies and finite 
localization lengths. These states have normalizable wave functions that 
vanish at infinite spatial region. Hence electrons in these states neither 
carry the bulk electric current nor contribute to electric conductances in 
the infinite system. In finite systems, however, localized states could 
behave like extended states if localization lengths are larger than or 
equal to the system size. They contribute to the electronic 
conductances if they have finite wave functions at current contact regions. 
Thus it is convenient to classify the one-particle energy 
region to several regime depending upon their localization lengths and the 
systems' sizes. There are three systems' sizes. One is the length of the 
system, $L_s$, the second one is the width in the 
Hall probe region, $L_{w1}$, and the third one is the width in the potential 
probe region, $L_{w2}$. Generally the first one is the largest, the second 
one is medium, and the third one is smallest. 

The localization length, $\lambda(E)$, depends on the localized state's 
energy and is known to behave as 
\begin{equation}
\lambda(E)=\lambda_0\vert E-E_l\vert^{-s},\ 
s\approx2,
\end{equation}
where $\lambda_0$ is a constant and $s$ is a critical exponent. 
The localization length diverges at the center of the Landau 
levels\cite{aoki} and becomes minimum at the middle between two Landau 
levels as is given in Fig.~25. Let $E_s$, $E_{w1}$, and 
$E_{w2}$ be the energy values where the localization lengths agree 
with the three system's sizes, 
$\lambda(E_s)=L_s$, 
$\lambda(E_{w1})=L_{w1}$, 
$\lambda(E_{w2})=L_{w2}$. Then one particle states in the energy range 
$\vert E-E_l\vert<E_s$ are extended states, 
and we call this energy region as the extended state region. 
One particle states in the energy range, $E_s<\vert E-E_l\vert<E_{w1}$, 
bridge from one edge to the other edge at the potential probe region and 
at the Hall probe region, 
and we call this energy region as the collapse regime. 
States in the energy range, $E_{w1}<\vert E-E_l\vert<E_{w2}$, 
bridge from one edge to the other edge at the potential probe region, 
and we call this energy region the dissipative quantum Hall regime (QHR). 
Finally states in the energy range, $E_{w2}<\vert E-E_l\vert$, are localized, 
and we call this energy region as the localized states region or QHR. 
Reasons why we use particular names for the last two regions will 
become clear later. 
The energy regions and typical wave functions in configuration space are 
written in Fig.~26. 
%
%
\begin{figure}[th]
\centerline{\includegraphics[width=7cm]{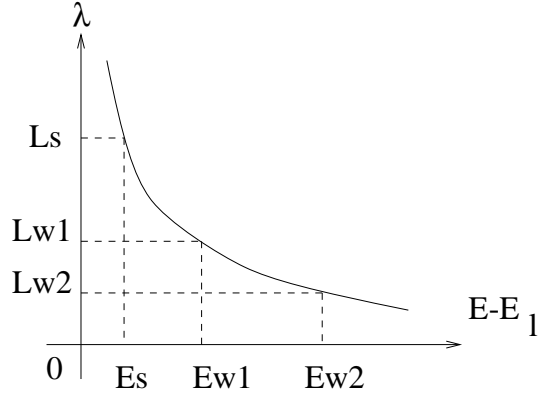}}
\vspace*{8pt}
\caption{A schematic illustration of localization length as a function of the 
one-particle energy measured from the center of Landau levels.  
}
\end{figure}

%
%
\begin{figure}[th]
\centerline{\includegraphics[width=7cm]{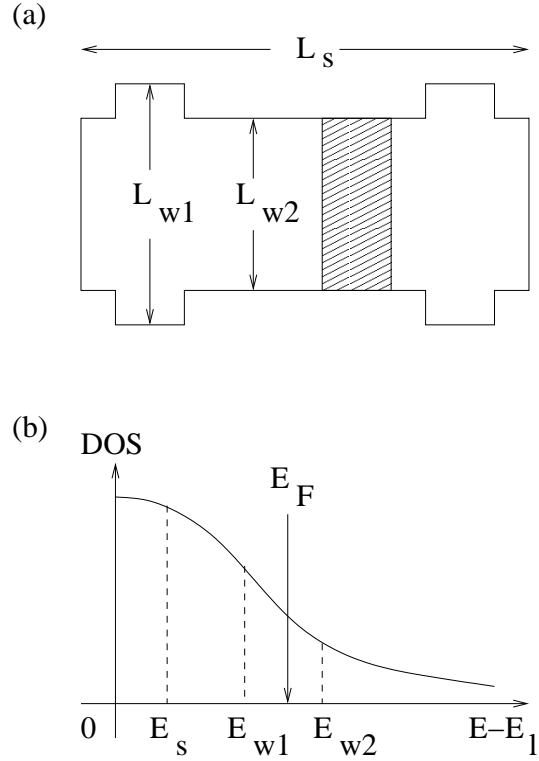}}
\vspace*{8pt}
\caption{(a) A schematic illustration of the two-dimensional electron system 
of a length $L_s$ and a width $L_{w1}$ in the Hall probe region and a width 
$L_{w2}$ in the potential probe region. (b) A schematic illustration of 
density of states as a function of the one-particle energy measured from 
the center of Landau levels.
}
\end{figure}

\subsection{Strip formation}

Non-uniform electron states where charge
density becomes high locally is formed by negative pressure effect. A strip 
of such non-uniform density region is formed also by localization 
effects. First one is due to interactions and the second one is due to 
disorders. One-particle wave 
functions are extended if its energy is in the center of Landau levels and 
cover whole spatial region from one end 
to another end in the system of no interactions. Interaction, however, 
modifies the many-body wave functions and compressible Hall gas of negative 
pressure is formed. Due to negative pressure, which is obtained in the 
previous section, this Hall gas shrinks. So the strip of compressible states 
of high density is formed. They are composed of one-particle states of the 
energies near the center of Landau levels. 

Second mechanism of forming charge non-uniform states is due to disorders in 
the finite systems. Wave functions of 
the energy near mobility edge have large localization length of the order of 
system sizes. These states which have the localization length in the region, 
$L_w<{ \lambda(E)}<L_s$, could bridge one edge to another edge in one 
direction of the sample, which is shorter in this direction than in the other 
direction but could not bridge in another direction of a sample. Consequently 
charge density of the localized states in this energy region becomes 
non-uniform and bridges only one sides. Their shape may not be simple but 
they behaves effectively like a strip of compressible gas. Thus the strip is 
formed in the extended states region near the mobility edge. 
The electric resistances of the strip states thus formed are studied in the 
next section. 

\section{Current activation from undercurrent: 
on precise determination of the fine structure constant }

In a normal metal the current flows through the extended one-particle states 
that are connected with the current contacts, and transport 
properties are determined by these extended states. 
The one-particle states that are disconnected from the current contacts 
should not affect transport properties. However in the quantum Hall system 
there are  extended states below the Fermi energy which carry the Hall 
current. Due to these states, one particle states which are disconnected 
from the current contacts are affected and contribute to the transport 
properties of the whole system in characteristic manners. 

We consider the situation where the electrons around the Fermi 
energy are in a strip which is disconnected with the current contacts and 
the extended states have the energy below the Fermi energy with a finite 
energy gap. Since extended states below the Fermi energy is connected with 
external current, the electrons below the Fermi energy carry the electric 
current. This  current dissipates no energy at low temperature. 
If current is induced in the states around the Fermi energy, which occurs by 
electron interaction, then the current carried by electrons dissipates 
energy, since there are many states around the Fermi energy. A small amount 
of current which is induced in the compressible region gives a 
small longitudinal resistance. Thus through higher order correlation effect, 
bulk Hall current activates the electrons around the Fermi energy which are 
disconnected from the current contacts. Consequently the current is induced 
in the strip of compressible states. We study physical effects of this 
induced current in the next part.

\subsection{Current activation from undercurrent as a new tunneling mechanism}

As shown in Sec.~IV, strip formed from localized electrons of localization 
lengths between the width, $L_w$, and length  of the 
system, $L_s$, ($L_w < L_s$), does not reach at least to one 
current contact. Hence injected current does not flow through the strip 
at zero temperature. The inject current flows through the electrons in 
lower Landau levels. A current in strip in this situation is induced by 
interactions and gives peculiar transport properties. Implications to the 
IQHE with finite injected currents at finite temperature will be discussed. 

%
%
\begin{figure}[th]
\centerline{\includegraphics[width=7cm]{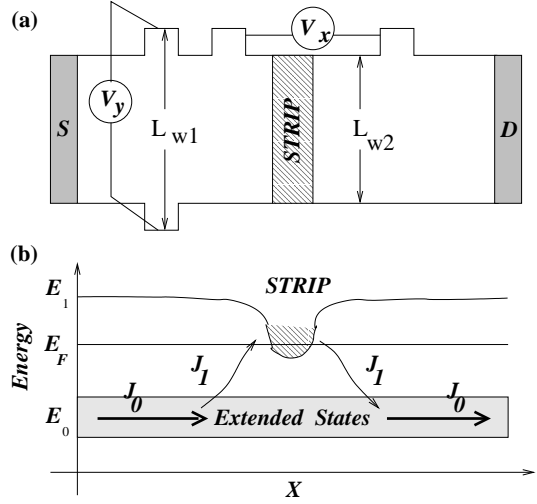}}
\vspace*{8pt}
\caption{(a) A schematic illustration of a potential probe and a Hall probe. 
A strip of compressible region bridges one edge to the other edge in the 
potential probe region of the width $L_{w2}$. S and D stand for the source 
and drain regions. (b) Electrons in the extended states which are in the lower 
Landau levels carry the current $J_0$. The electrons in the strip are 
around the Fermi energy and absorb the current $J_1$ from the lower levels. 
}
\end{figure}

We study in the following the situation where several lower Landau levels 
are filled completely and one Landau level is partially filled as shown in 
Fig.~28. 
The strip is formed in the highest occupied Landau level and 
bridges one edge to the other edge. 
Although Fermi energy is in the strip state energy region, 
the strip is disconnected from the current contacts. 
Injected electric current flows through the lower Landau levels, 
and does not flow through the isolated compressible states of higher Landau 
levels if there is no interaction. 
In reality the interaction exists and modifies the isolated states. 
At finite temperature the current is activated into the isolated strip 
states from lower Landau levels as shown in Fig.~28. 
We study the effects of Coulomb interaction at finite temperature 
and find that the strip states have small amount of 
electric current at low temperature. 
The magnitude of induced current depends on injected current as well 
as on temperature. 

From Fermi-Dirac statistics, one electron fills one state and states 
are filled from lower energy levels to higher energy levels. Among electrons
in the many-body states, only the electrons near the Fermi surface make 
transitions  with low energy perturbations. The dissipation of the energy 
occurs only from the states near the Fermi 
surface and does not occur at the lower energy levels. In the quantum Hall 
system, the bulk current is carried by the electrons in lower levels and in 
the levels near Fermi surface. So in the present situation shown in Fig.~28, 
dissipationless current flows in the lower levels and induced current in 
the strip in the middle of sample near Fermi surface dissipates energy. 

%
%
\begin{figure}[th]
\centerline{\includegraphics[width=5cm]{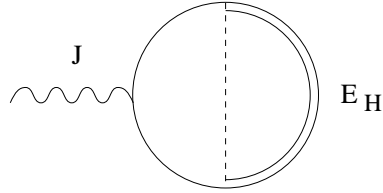}}
\vspace*{8pt}
\caption{Feynman diagram which gives an induced current. The solid line 
stands for the electron propagator in the strip state and the double line 
stands for the electron propagator in the lower Landau levels. The dashed 
line shows the Coulomb interaction.
}
\end{figure}

The second order correction to the current correlation functions from the 
Coulomb interaction is expressed in Feynman diagram of Fig.~29 
and gives the induced current to the strip states. One electron line 
(double line) stands for those of lower Landau levels which are connected 
with current contacts and are carrying current. Thus these states depend 
on the Hall electric field.\cite{br1,br2} 
The other line (solid line) stands for those of strip states around Fermi 
energy which are disconnected from current contacts and are not carrying 
current in the lowest order. The dashed line stands for the Coulomb 
interaction. To compute induced electric current, the current operator is 
inserted in the strip state line and is shown as wavy line in Fig.~29. 
We express operators, propagators, vertices, and others in the current 
basis. The induced current is calculated as
\begin{eqnarray}
j_{\rm ind}^x&=&{1\over\beta}\sum_{\omega=i\omega_n}
\int{d^2 p d^2k\over(2\pi)^4}{\rm Tr}\left(
 V({\bf k}){\tilde S}^{(c)}(p-ak)\right.\nonumber\\
&&\left.
\times{\tilde \Gamma}^x(p-ak,p-ak){\tilde S}^{(c)}(p-ak){\tilde S}(p;
E_H)\right)\\
&=&{1\over\beta}\sum_{\omega=i\omega_n}
\int{d^2 p d^2k\over(2\pi)^4} V({\bf k})
\vert f_{l,l+1}({\bf k})\vert^2 \nonumber\\
&&\times\partial_x S_{l,l}(p;E_H)S^{(c)}_{l+1,l+1}(p-ak),\nonumber
\end{eqnarray}
where $S(p;E_H)$ is the propagator of the electrons of the energy determined 
with the Hall electric field, $E_l+eE_H^{\rm eff}{a\over2\pi}p_x$ which 
depends on the Hall electric field 
$E_H$ and $S^{(c)}(p)$ is the propagator for the stripe states, Eq.~(110), 
which does not depend on $E_H$. Actually for the propagator $S^{(c)}(p)$ 
in the strip due to disorders, we use the same form as the propagator in 
the strip due to interactions for simplicity. 
$\omega_n=(2n+1)\pi/\beta$ is the Matsubara frequencies. 
The chemical potential $\mu$ is in between $E_l$ and $E_{l+1}$. 
In the above calculation, we considered only the mixing term between the 
Landau levels $E_l$ and $E_{l+1}$ which is the dominant term in the 
induced current. The asymmetry parameter is fixed at $r_s=1$ for simplicity. 
Then the induced current reads 
\begin{eqnarray}
j_{\rm ind}&=&\int{d^2 p d^2k\over(2\pi)^4}V ({\bf k})
\vert f_{l,l+1}({\bf k})\vert^2 v_x\beta 
e^{\beta\Delta E_1({\bf p}-a{\bf k})}\theta(-\Delta 
E_2({\bf p})),\nonumber\\
\Delta E_1({\bf p})&=&E_l+eE_H^{\rm eff}{a\over2\pi}p_x-\mu,
\label{eh}\\
\Delta E_2({\bf p})&=&E_{l+1}+\varepsilon_{l+1}({\bf p})-\mu,\nonumber\\
v_x&=&eE_H^{\rm eff}{a\over2\pi},\nonumber
\end{eqnarray}
where $\Delta E_1({\bf p})$ comes from the effective band width of the 
electrons in the Hall electric field. $E_H^{\rm eff}$ is an effective 
Hall electric field and is given as $E_H^{\rm eff}=E_H/\gamma$, which is 
enhanced by the strong localization effect.\cite{br1,br2} 
In experiments,\cite{kawa} $1/\gamma$ is on the order of 10.  
$\Delta E_2({\bf p})$ comes from the band electrons which is computed 
in the previous section. 

The induced current is given by,
\begin{eqnarray}
j_{\rm ind}&=&V_l e^{\beta(E_l-\mu)}\sinh{\beta e E_H^{\rm eff}\over 2},
\label{act}\\
V_l&=&{4\pi\nu'q^2\pi^{3/2}\over a^2 (l+1)}\int_0^\infty dx x^{1/2}\{L_l
^{(1)}(x)\}^2 e^{-x}.
\nonumber
\end{eqnarray} 
The value depends on the Hall electric field, the total current, and the 
temperature. The temperature dependence is of activation type and the gap 
energy is linear in $E_H$. 

\subsection{Induced current in a strip of compressible Hall gas}

The temperature dependence of the current is of activation type. 
At relatively high temperature, the electron temperature is the same as 
that of the whole system. At low temperature, however, the band electron 
temperature is different from the system's temperature. Electron temperature 
is determined from the energy due to the electric resistance and the heat 
conduction from the electron system to the lattice system. The temperature 
determines the electric resistance and conversely the electric resistance 
determines the temperature. Hence they are determined 
self-consistently.\cite{komiyama} 

The induced current of the strip depends on the total injected current 
through the Hall electric field. Effective band width Eq.~(\ref{eh}) of 
extended Landau levels in the system of Hall electric field increases with 
Hall electric field. This agrees with the experiments of the breakdown of 
the IQHE due to the injected current, where the activation energy 
depends on the injected current. So we use these previous results in the 
present work. 

\subsection{Longitudinal resistance}

The total longitudinal resistance is the ratio between 
the induced electric field in the compressible gas region and the 
current density and is given as, 
\begin{eqnarray}
R_{xx}&=&{ R^{(s)}_{xx} ej_{\rm ind} \over \sigma_{xy}E_H},     \\
\label{rho}
R^{(s)}_{xx}&=&\left({ e^2 \over h}\right)^{-1}
\end{eqnarray}
where $R^{(s)}_{xx}$ is the longitudinal resistance of the stripe state. 
The particular dependence of the longitudinal resistance on the Hall field 
is obtained for the first time. This result are compared with the experiment 
of Kawaji et al.\cite{kawa} in Fig.~30. 
%
%
\begin{figure}[th]
\centerline{\includegraphics[width=7cm]{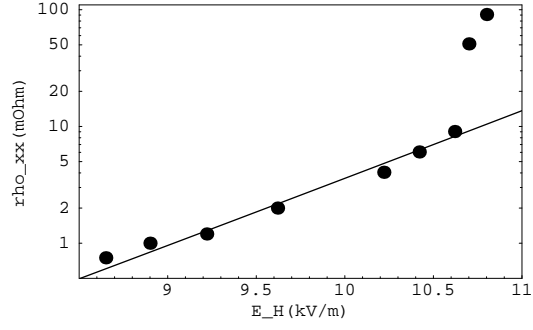}}
\vspace*{8pt}
\caption{Comparison of the theoretical formula of induced longitudinal 
resistivity with the experiment of Kawaji et al\cite{kawa}. The solid 
line stands for the theoretical curve and the circles correspond the 
experiments.
}
\end{figure}

In the experiment, the filling factor is 4, magnetic field is 5.7 T, and 
the temperature is 0.75 K. The electron 
temperature could be different from the above value due to heating effect and 
is not known experimentally. So we fit the theoretical curve by changing the 
temperature, $\gamma$, and $\nu'$. As shown in Fig.~30, reasonable agreements 
are obtained when the temperature is 1.576 K, $\gamma=0.065$, $\nu'=0.5$ 
and $\mu=E_l+\hbar\omega_c/2$. Actually the temperature and the $\gamma$ 
are almost the same even though  $\nu'$ is varied substantially. 
This temperature is that of the electron 
system at the strip and is different from the experimental value which is at 
the lattice system. The large deviation at large Hall electric fields is 
caused by the breakdown of IQHE, which will be discussed later. 

\subsection{Various regimes with current increases}

The transport properties of the finite system depend on the system sizes, 
the Fermi energy, and the magnitude of the induced current. 
If the width in the Hall probe region is wider than the width in the 
potential probe region, longitudinal resistance and Hall resistance behave 
very differently. 

As the current increases, band widths become broad from Eq.~(\ref{eh}). 
Boundary between the extended states and localized states, mobility edges, 
move and localization length of given energy increase. 
Hence depending on the magnitude of the current,current carrying states 
occupy the whole area differently as is shown in Fig.~31. In {\bf QHR}, 
current carrying states become small islands and are isolated. At a larger 
current, the second region, {\bf dissipative QHR}, where the current 
carrying states become larger and bridge the sample in the potential probe 
region appear. At a more larger current, the third region, 
{\bf post collapse regime},
where the current carrying states become larger and bridge the sample in 
the potential probe region and in the Hall probe region appear. If the 
current becomes even larger,whole area of the sample are covered by the 
current carrying states and QHE does not occur. 
QHR is {\bf breakdown}. 
The resistances show several different behaviors. 
%
%
\begin{figure}[th]
\centerline{\includegraphics[width=7cm]{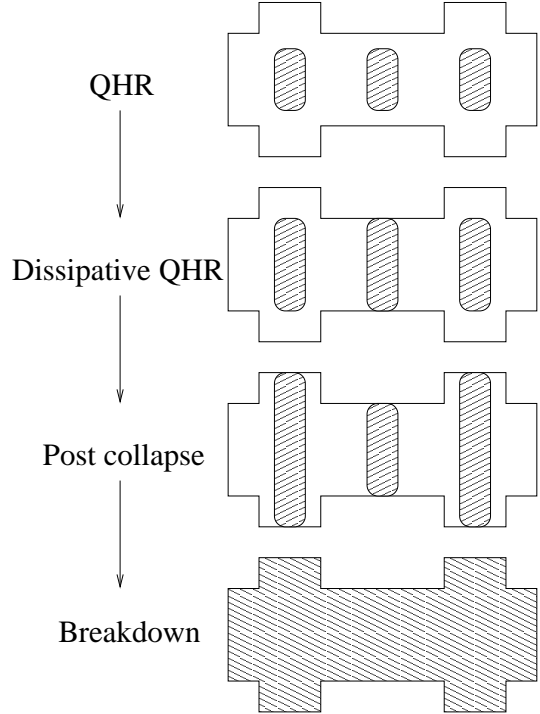}}
\vspace*{8pt}
\caption{ A schematic illustration of current carrying states. Shaded 
regions show the current carrying states and become larger as the 
current is increased. 
}
\end{figure}

{\bf Dissipative QHR}

When the localization length of a state reaches the 
width of only the potential probe region, this state contributes to 
longitudinal resistance but not to the Hall resistance. 
Potential probe region has a temperature dependent electric resistance of 
activation type as is given in Eq.~(123) and Eq.~(124). 
In the Hall probe region, all the particle states around the Fermi energy are 
still  localized states which have shorter localization lengths than the 
width. Hence the Hall probe region is considered as the QHR. 

The Hall resistance measured in the wider Hall probe region is computed by 
the topological invariant expression Eq.~(\ref{topological}) with the 
momentum in the finite system. The momentum becomes discrete in the finite 
system.
As was shown in Ref.~\cite{br1}, this topological invariant is determined 
by the magnetic field and has no dependence on spatial component of the 
momentum. Hence this topological invariant does not change the value even 
when the integration variables are replaced with the discrete values. 

Thus the quantized Hall resistance has no finite 
size corrections in the localized state region.\cite{br1} Hence the Hall 
resistance measured in the present situation agrees with the exactly 
quantized value. Thus resistivities at low temperature are given in our 
theory as,
\begin{eqnarray}
\rho_{xx}&=&\tilde\rho e^{-\beta\Delta E_{\rm gap}},\nonumber\\
\rho_{xy}&=&\left({e^2\over h}N\right)^{-1},\label{diss}\\
\Delta E_{\rm gap}&=&\mu-E_l-{ eE_H^{\rm eff}a \over 2},\nonumber
\end{eqnarray}
where $\tilde\rho$ is temperature independent. The finite temperature 
correction to $\rho_{xy}$ was estimated before. The correction becomes 
independent from $E_H$ and is negligible in the current situation. 
Thus $\rho_{xy}$ is quantized even though $\rho_{xx}$ does not vanish. 

Equation (\ref{diss}), show that the Hall resistance is quantized 
even though the longitudinal resistivity does not vanish. This is a new 
regime of IQHE, which has not been expected from the naive picture of IQHE. 
Equation (\ref{diss}) shows that the new dissipative QHR is realized only in 
the finite system.  If the width in the Hall probe region is the same as the 
width in the potential probe region, the dissipative QHR does not exist. 

\subsection{Collapse and breakdown of IQHE}

%
%
\begin{figure}[h]
\centerline{\includegraphics[width=7cm]{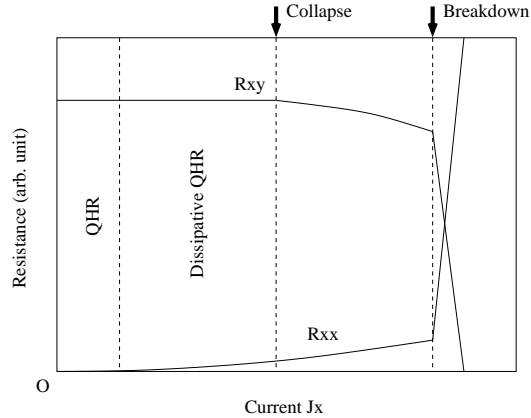}}
\vspace*{8pt}
\caption{A schematic illustration of the Hall resistance and the 
longitudinal resistance in the QHR, the dissipative QHR, the  collapse, 
and the breakdown region as the inject current increases. }
\end{figure}
If the current increases further, the extended state regions broaden 
in energy by the amount that is proportional to Hall electric field. 
Mobility edges move with Hall electric field outward from the center 
of Landau levels. The localization lengths of the localized states become 
larger. At some current value, they reach the spatial width of the Hall 
probe region. Wave functions in this energy region bridge both of the 
potential probe region and the Hall probe region. The electric current in 
this region is given by the previous form of the activation type in 
Eq.~(\ref{act}). If the Fermi energy is in this region, not only Hall 
resistance has a small correction but also longitudinal resistances becomes 
finite at low temperature. They are given in our theory as,
\begin{eqnarray}
\rho_{xx}&=&\tilde\rho e^{-\beta\Delta E_{\rm gap}},\\
\Delta\rho_{xy}&=&{\tilde \rho}'e^{-\beta\Delta E_{\rm gap}},
\end{eqnarray}
where $\tilde\rho$ and $\tilde\rho'$ are independent from the temperature. 
This corresponds to the collapse of IQHE. 

When the current increases further, 
the localization lengths of the localized states become even larger and 
reach the spatial length of the system. Wave functions in this energy region 
cover whole area of the system and reach the current contacts. 
If the Fermi energy is in this region, the Hall resistance and the 
longitudinal resistances have finite corrections and are given by, 
\begin{eqnarray}
\rho_{xx}&=&{h\over e^2}\delta,\\
\Delta\rho_{xy}&=&{h\over e^2}\delta',
\end{eqnarray}
where $\delta$ and $\delta'$ are finite numbers which do not vanish 
in $\beta\rightarrow\infty$. This corresponds to the breakdown of the IQHE 
in the finite system. If the current increases further and exceeds a critical 
value in which the energy gap $\Delta E_{gap}$ vanishes completely, 
the localization length becomes infinite, the longitudinal resistance 
becomes finite, and Hall resistance deviates from a quantized value 
substantially. The IQHE in the infinite system disappears then. The critical 
value of Hall electric field, $E_c$, is given from 
Eq.~(\ref{eh}) as,
\begin{equation}
{\hbar\omega_c\over2}=eE_c^{\rm eff} a. 
\end{equation}
The critical Hall electric field is proportional to,
\begin{equation}
E_c={\hbar\omega_c\over2e}{\gamma\over a}.
\end{equation}
$E_c$ is proportional to $B^{3/2}$ which is consistent with the 
experiment.\cite{kawa}

\subsection{Precise determination of the fine structure constant from IQHE}

The unusual properties found in Section 3 lead the systems to have several 
interesting transport properties depending upon the magnitude of several 
parameters such as current, system sizes, and others. One of the critical 
features is that the quantum Hall gas has a tendency to shrink. Hence the 
electric property of some spatial region could be confined in that region 
and does not expand to the whole system. From Section 5.4, even though an 
energy dissipation occurs in some region of the sample, the other part of 
the same sample could give the exactly quantized value of the Hall 
conductance.  This happens if the other part of the sample is completely 
occupied by the localized electrons. The Hall conductance measured in this 
part of the sample is quantized exactly. This is an important 
result concerning metrology of the quantum Hall effects and is a key factor 
for the precise determination of the fine structure constant.    
In fact there is a  small energy dissipation in the real measurement 
from the current contact area because there is a potential drop in the 
direction parallel to the current. However it is possible to measure 
the fine structure constant precisely from  the IQHE.

\section{ \bf Summary}

Fundamental physical constants such as the light velocity,the electron 
charge,and the Plank constant  played important roles in establishing 
modern physics. They are connected with relativity, electromagnetism, and 
microscopic world of matters. It is important to measure these constants
precisely to understand the law of nature. The fine structure constant, 
${\alpha}$ is the unique parameter which is a combination of these constants 
and is the expansion parameter of QED. Computations of higher order 
corrections and  the experimental measurements of QED with great precisions 
have been achieved. Comparisons of the theory with the experiments 
could supply  a useful  test of QED and standard model of high energy 
physics. Since the most accurate value of  ${\alpha}$ is obtained from the 
QHE, it is necessary to know if the precise value of ${\alpha}$ is known 
from the QHE. To find out this problem, we formulate field theory of the 
QHE in such manner that makes derivation of rigorous field theoretical 
identities transparent. The von Neumann lattice representation of guiding 
center variables is suitable representation for developing field theory. 
Based on this representation, we give the proof of the integer quantization 
of the Hall conductance in the realistic situation and studied quantum Hall 
gas. Owing to Ward-Takahashi identity between the propagator and the 
vertex part, the Hall conductance is written by the special topological 
invariant of the mapping in three dimensional space defined by the 
propagator in the Landau levels.  

The von Neumann lattice 
representation is useful also for studying anisotropic quantum Hall gas 
phase. The ground state, the physical properties, the orientation of the 
anisotropic quantum Hall gas under the small external 
perturbation, and the excited states are studied. The translational and 
rotational symmetries  are  clarified and the broken phase is studied. 
Implications of the quantum Hall gas states to the metrology of the 
QHE are also studied at the end. 

Gauge invariance and magnetic field are two important ingredients for 
the topological invariant expression of Hall conductance of section II. 
We may ask ourself ``Does the same topological invariant appear in physical 
systems without magnetic field? '' This problem has been studied  
and affirmative answer has been obtained. Namely if time reversal is 
broken, the same topological invariants appear. One  simple system is a 
free massive Dirac field\cite{ishikawa,topological} in planer space and 
the other is time reversal violating condensed matter system such as unusual 
superconductor\cite{goryo ishikawa} and others\cite{volovik}.  
In the first case, the spectrum asymmetry of Dirac operator 
which causes chiral anomaly and fractional charge in even dimensional 
space-time is the origin. The mass 
term breaks time reversal invariance, and U(1) symmetry is 
preserved. Hence the value of the topological invariant is  quantized 
without any  correction. On the other hand, in time reversal 
violating superconductor, U(1) symmetry is broken spontaneously. So 
topological invariant is not necessarily quantized and has a correction. 

Other phases have been seen recently but are not discussed in the 
present work. They include a series of FQHS 
around half filling, pairing state at ${\nu=5/2}$ \cite{5/2pairing}, 
reentrant charge density wave \cite{nbn}, and
many others. The von Neumann lattice representation would be useful for 
studying these states as well. 

\section*{Acknowledgements}

We thank J. L. Birman, G. H. Chen, K. Cooper,  A. T. Dorsey, 
J. Eisenstein, A. Endo, J. Goryo, Y. Hosotani, Y. Iye, R. Jackiw, S. Kawaji, 
R. M. Lewis, T. Ochiai, W. Pan, K. Tadaki, L. Tevlin, P. Wiegmann,
Y. S. Wu, K. Yang, and J. Zhu for useful discussions. 
This work was partially 
supported by the special Grant-in-Aid for Promotion of Education and 
Science in Hokkaido University provided by the Ministry of 
Education, Culture, Sports, Science, and Technology, Japan and 
by the Grant-in-Aid for Scientific Research on Priority area (Dynamics of 
Supertrings and Field Theories) (Grant No.13135201), Ministry of 
Education, Culture, Sports, Science, and Technology, Japan, 
Clark Foundation,
and Nukazawa Science Foundation.

\end{document}